\documentclass[12pt]{iopart}

\expandafter\let\csname equation*\endcsname\relax
\expandafter\let\csname endequation*\endcsname\relax
\RequirePackage{amsmath}

\usepackage[english]{babel}
\usepackage{bbold}
\usepackage{upgreek}
\usepackage{dsfont}
\usepackage{graphicx}
\usepackage{comment}
\usepackage{braket} 
\usepackage{colortbl}
\usepackage{cite}

\newcommand{\bls}[1]{\mathbf{#1}} 
\newcommand{\blg}[1]{\boldsymbol{#1}} 
\newcommand{\tcr}[1]{\textcolor{black}{#1}}

\begin{document}

\title{Electric trapping and circuit cooling of charged nanorotors}

\author{Lukas Martinetz$^{1}$, Klaus Hornberger$^{1}$, and Benjamin A. Stickler$^{1,2}$} 

\address{$^{1}$University of Duisburg-Essen, Faculty of Physics, Lotharstra\ss e 1, 47048 Duisburg, Germany}
\address{$^{2}$Imperial College London, Quantum Optics and Laser Science, Exhibition Road, London SW72AZ, United Kingdom} 

\begin{abstract}
The motion of charged particles can be interfaced with electric circuitry via the current induced in nearby pick-up electrodes. Here we show how the rotational and translational dynamics of levitated objects with arbitrary charge distributions can be coupled to a circuit and how the latter acts back on the particle motion.
The ensuing cooling rates in series and parallel RLC circuits are determined, demonstrating that quadrupole ion traps are well suited for implementing all-electric cooling. We derive the effective macromotion potential for general trap geometries and illustrate how consecutive rotational and translational \tcr{resistive} cooling of a microscale particle can be achieved in linear Paul traps.
\end{abstract}

\maketitle

\section{Introduction}

Cooling and controlling the rotational and translational motion of nano- and microscale levitated particles is crucial for force and torque sensors \cite{ranjit2015attonewton,kuhn2017b,ahn2020ultrasensitive,hempston2017} and for future quantum superposition tests \cite{arndt2014a,millen2020quantum}. Optical techniques are well established \cite{millen2020optomechanics,gieseler2021optical}, enabling  cooling the center-of-mass motion of spherical objects into the quantum regime \cite{delic2020cooling,tebbenjohanns2020motional,magrini2020optimal} and rotational control of aspherical particles \cite{kuhn2017a,ahn2018optically,ahn2020ultrasensitive,reimann2018ghz,van2020optically,rashid2018precession,monteiro2018optical,bang2020,vanderlaan2020observation,stickler2021}. However, optical methods are limited by the unavoidable impact of photon scattering \cite{jain2016direct,vanderlaan2020observation} and absorption \cite{millen2014nanoscale,hebestreit2018measuring}, and optical traps may become  unstable at low pressures \cite{kiesel2013cavity,millen2014nanoscale,vovrosh2017parametric}. 

An  alternative is to electrically \tcr{\cite{millen2015cavity,alda2016trapping,delord2017a,goldwater2018levitated,bykov2019,delord2020spincoupling,dania2020} \tcr{or magnetically} \cite{cirio2012quantum,hsu2016cooling,rusconi2017quantum,prat2017ultrasensitive,o2019magneto,timberlake2019acceleration,latorre2020chip,gieseler2020single}} levitate \tcr{charged or magnetized} particles, \tcr{which can be controlled} without lasers. Quadrupole ion traps are well established in the field of levitated optomechanics \cite{millen2015cavity,delord2017a,delord2020spincoupling,dania2020}, providing stable levitation for  a wide range of particle masses \cite{alda2016trapping,goldwater2018levitated,bykov2019}. Such Paul traps have been combined with optical fields \cite{delord2018,delord2020spincoupling} and cavities \cite{millen2015cavity,fonseca2016} to manipulate and cool the particle motion. Electric feedback cooling via external electrodes has been realized in both electrical \cite{dania2020} and optical \cite{tebbenjohanns2019cold,conangla2019optimal}  setups. 

Also all-electric schemes are feasible, given that purely electric cooling \tcr{\cite{itano1995, dehmelt1968bolometric,church1969radiative,wineland1975principles, brown1986,major2005charged,kaltenbacher2011resistive,di2015toward,cornell1989single}} and manipulation techniques devised for atomic ions \cite{kotler2017,tian2004,kielpinski2012} can be applied to charged objects of arbitrary shape and size. However, the particle dynamics are then  determined not only by the electric monopole, but depend also on higher multipole moments, which couple the rotational and translational particle motion to the time-dependent trapping and manipulation fields \tcr{\cite{joseph2009long,hashemloo2015rotational,delord2017a,martinetz2020,rudyi2021time}}.

In the present article, we provide the theory required for interfacing the ro-translational motion of trapped nanoparticles with electric circuitry by means of dedicated pick-up electrodes. Based on this, we show how the particle motion can be cooled selectively via the energy dissipation in a resistor. 
This may facilitate all-electric 
precision experiments with charged levitated particles, such as mapping the electrostatic interaction with nearby surfaces \cite{winstone2018}, probing the existence of milli-charged particles \cite{moore2014,rider2016}, detecting deviations of Coulomb's law at short distances \cite{moore2020searching},
or entangling the particle motion with a superconducting qubit \cite{martinetz2020}.
Such schemes depend crucially on how the combined rotational and translational motion of the nanoparticle is affected by the electric fields and how it acts back on the circuit.

The structure of this article is as follows. In Sect.~\ref{sec:coupling}
we derive the coupled  equations of motion for the trapped nanoparticle and the electric circuit connecting the pick-up electrodes,  accounting for arbitrary charge distributions, particle shapes, and electrode configurations. We then argue in Sect.~\ref{main_resistive_cooling} that resistive cooling of the rotational and translational particle motion is feasible in state-of-the-art setups by providing analytic expressions for the damping rates in parallel and series RLC circuits. Section \ref{sec_electric_levitation} shows how the effective potential for the combined rotational and translational macromotion in a quadrupole ion trap is obtained by separating the small-amplitude micromotion from the slow macromotion in the driving field. The resulting effective dynamics are found to be in excellent agreement with numerically exact  simulations. Section~\ref{eff_pot_special_particle} specifies the macromotion potential for the most common trap geometries and particle shapes, allowing us to discuss in Sect.~\ref{sec:equilibration} their equilibration dynamics in presence of friction and diffusion due to a gas. We finally show in Sect.~\ref{sec:exp} by numerical simulation  that individual  degrees of freedom of a  microscale particle can be cooled consecutively by tuning the circuit parameters, before presenting our conclusions in Sect.~\ref{sec:conclusions}. 

\section{Particle-circuit coupling}
\label{sec:coupling}
We consider a nanoparticle of mass $m$, with moments of inertia $I_i$, and an arbitrary surface charge distribution, coupled to an electric circuit, see Fig.~1. The inertia tensor ${\rm I}(\Omega) =\sum_{i} I_i {\bf N}_i(\Omega)\otimes {\bf N}_i(\Omega)$ depends  on  the  nanoparticle  orientation $\Omega$ through  the principal axes ${\bf N}_i(\Omega) ={\sf R}(\Omega)\,{\bf e}_i$, obtained by rotating the space-fixed coordinate axes ${\bf e}_i$ with the rotation tensor ${\sf R}(\Omega)$, see \ref{appendix_hamiltonian}.  The  charge  distribution  on  the  particle  surface is  characterized  by  the  total  charge $q$, by the  orientation-dependent dipole vector $\blg{p}(\Omega) =\sum_i p_i {\bf N}_i(\Omega)$ with dipole moments $p_i$, by the quadrupole tensor ${\sf Q}(\Omega)= \sum_{ij} Q_{ij}{\bf N}_i(\Omega)\otimes {\bf N}_j(\Omega)$ with  quadrupole  moments $Q_{ij}$,  as well as  by higher  multipole  moments.   All  moments  are  defined  with  respect to  the  particle  center  of  mass  and  are  constant  in  the  body-fixed  frame.   The  particle  moves and  revolves  in  the  time-dependent  trapping  potential $V_{\rm tr}({\bf R},\Omega,t)$, where ${\bf R}$ denotes the center-of-mass position.  This potential will be derived in  Sec.~\ref{sec_electric_levitation} for the case of electrical quadrupole  traps,  but  it could  also  be  due to optical or magnetic levitation fields.

\begin{figure*}[t]
\centering
\includegraphics[width = 1\textwidth]{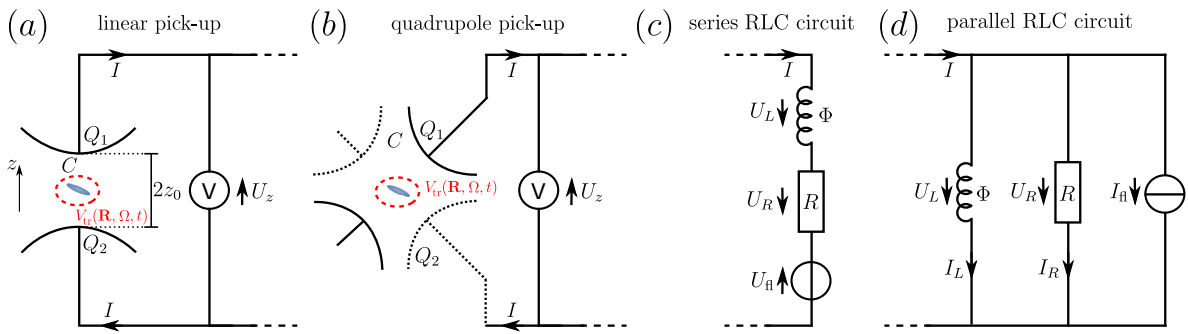}
\caption{The motion of the trapped nanoparticle induces a current in the pick-up electrodes, which can for instance be arranged in a linear (a) or a quadrupole configuration (b). The electrodes can be connected either to a series (c) or to a parallel (d) RLC circuit for cooling the rotational and translational particle dynamics. The arrows indicate the direction of increasing
electrostatic potential for positive $U$ and the direction of the electron flux for positive $I$. The voltage and current fluctuations due to Johnson-Nyquist noise in the resistor are indicated by $U_{\rm fl}$ and $I_{\rm fl}$. } \label{figure1}
\end{figure*}

The nanoparticle dynamics can be interfaced with electric circuitry by placing two pick-up electrodes close to the trapping region, see Fig.~\ref{figure1}. The two electrodes thus form a capacitor of capacitance $C$. The charge offset $Q=(Q_1-Q_2)/2$ between the pick-up electrodes depends on both the voltage offset $U_z$ and on the particle position and orientation due to the induced charge. The presence of additional electrodes can also contribute, as described by a capacitance matrix. Since this contribution vanishes in the symmetric trap setups considered below, we neglect it in the following.

We denote by $U_z \Phi_0({\bf r})$ the electrostatic potential in the trapping region that would be present without the particle if the pick-up electrodes are at $\pm U_z/2$ and all other electrodes  grounded. The exact form of $\Phi_0({\bf r})$ will be specified below in Sect.~\ref{main_resistive_cooling} for the linear and quadrupole pick-up configurations shown in Fig.~\ref{figure1}(a), (b). This potential allows  calculating (i) the charge offset between the pick-up electrodes in presence of the particle and (ii) the electric force and torque acting on the particle at a given voltage offset $U_z$.

First, the charge offset follows from Green's reciprocity theorem \cite{martinetz2020}\tcr{, assuming that all retardation effects can be neglected,} as
\begin{equation}\label{sec3_coupling}
Q=-Q_{\rm ind}(\bls{R},\Omega)+CU_z,
\end{equation}
where the induced charge is determined by position- and orientation-dependent charge density on the nanoparticle $\rho({\bf r};{\bf R},\Omega)$, 
\begin{align}\label{eq:indcharge}
    Q_{\rm ind}(\bls{R},\Omega)=&\int d^3{\bf r}\, \rho(\bls{r};\bls{R},\Omega)\Phi_0(\bls{r}).
\end{align}
Using that the nanoparticle charge density can be transformed into the body-fixed frame as $\rho({\bf r}; {\bf R},\Omega) =\rho_0[\mathsf{R}^{-1}(\Omega)(\bls{r}-\bls{R})]$, yields
\begin{align}
    Q_{\rm ind}(\bls{R},\Omega)=&\int d^3{\bf r}' \rho_0(\bls{r}')\Phi_0(\bls{R} + \mathsf{R}(\Omega)\,\bls{r}').
\end{align}

Second, the total force experienced by the nanoparticle can be separated into three contributions: (a) The force ${\bf F}_{\rm im}(\bls{R},\Omega)$ due to image charges in nearby metal surfaces, that is the force if all electrodes are grounded. (b) The trapping force ${\bf F}_{\rm tr}(\bls{R},\Omega,t)$ with corresponding potential $V_{\rm tr}(\bls{R},\Omega,t)$. (c) The force due to a voltage between the pick-up electrodes. The latter follows from integrating the force density acting on the particle charge distribution in an inhomogeneous electric field $-\rho(\bls{r}; {\bf R},\Omega)U_z\partial\Phi_0/\partial \bls{r}$. The equation of motion for the center-of-mass momentum $\bls{P}$ can thus be expressed in terms of the induced charge $Q_{\rm ind}$,
\begin{subequations}\label{main_force_and_torque}
\begin{equation}\label{main_force}
\frac{d}{dt}{\bls{P}}=  {\bf F}_{\rm im} +\bls{F}_{\rm tr}  -U_z \frac{\partial }{\partial \bls{R}}Q_{\rm ind}.
\end{equation}

Similarly, the angular momentum vector $\bls{J}$ of the particle is subject to
\begin{equation}
\label{main_torque}
\frac{d}{dt}{\bls{J}}= {\bf N}_{\rm im} + \bls{N}_{\rm tr}  -U_z\bls{T},
\end{equation}
\end{subequations}
where  $\bls{N}_{\rm im}(\bls{R},\Omega)$ and $\bls{N}_{\rm tr}(\bls{R},\Omega,t)$ are the torques due to the image charges and the trap potential, respectively, and 
\begin{equation}\label{eq:unittorque}
    \bls{T}(\bls{R},\Omega)=\int d^3{\bf r}' \rho_0(\bls{r}')\big[\mathsf{R}(\Omega)\bls{r}'\big]\times\frac{\partial}{\partial {\bf R}}\Phi_0\big[\bls{R}+\mathsf{R}(\Omega)\,\bls{r}'\big].
\end{equation}

The image forces and torques can be readily given for a particle close to the center of a flat plate capacitor (see \ref{appendix_image_potential})
\begin{subequations}\label{image_force_torque}
\begin{align}
    {\bf F}_{\rm im}(\bls{R},\Omega) =&\;\frac{7q \zeta(3) }{32\pi \varepsilon_0 z_0^3} \big[{\bf e}_z \cdot ( q {\bf R} + {\blg p} )\big] \,{\bf e}_z,
\\
    {\bf N}_{\rm im}(\bls{R},\Omega) = &\; \frac{7q\zeta(3)}{32\pi \varepsilon_0 z_0^3} \Big \{ ({\bf e}_z \cdot {\bf R}) \,{\blg p}
    + \frac{5}{14q} \big({\bf e}_z \cdot {\blg p}\big){\blg p} + \frac{3}{14} {\sf Q}\, {\bf e}_z \Big \} \times {\bf e}_z,
\end{align}
\end{subequations}
with the Riemann $\zeta$-function $\zeta(3)\approx 1.20$.

The particle dynamics are fully determined by Newton's equations (\ref{main_force}) and (\ref{main_torque}), complemented by the kinematic laws $\dot{\bls{R}}=\bls{P}/m$ and $\dot{\mathsf{R}}=\left({\rm I}^{-1}\bls{J}\right)\times\mathsf{R}$. 
In case of a vanishing voltage offset, $U_z=0$, the particle motion decouples from the RLC resonator.  The remaining forces and torques are then conservative, so that the motion in absence of the the circuit degrees of freedom is given by the nanoparticle  Hamiltonian
\begin{align}\label{eq:npham}
    H_{\rm np} = & \frac{1}{2}\bls{J}\cdot{\rm I}^{-1}(\Omega)\,\bls{J}+\frac{\bls{P}^2}{2m}+V_{\rm tr}+V_{\rm im}.
\end{align}
If Euler angles are used to specify the particle orientation $\Omega$ in (\ref{eq:npham}), the angular momentum vector $\bls{J}$ in the kinetic energy is expressed by the corresponding canonically conjugate momenta, see \ref{appendix_hamiltonian}. 

The total  potential energy contains the trapping potential $V_{\rm tr}(\bls{R},\Omega,t)$ and the contribution $V_{\rm im}(\bls{R},\Omega)$ of  the image charges. Indeed, one finds that the image force and torque in Eq.~(\ref{image_force_torque}) can be derived from
\begin{align}\label{image_potential}
    V_{\rm im}({\bf R},\Omega) = & - \frac{\zeta(3)}{64\pi \varepsilon_0 z_0^3} \left [ 7q^2 ({\bf e}_z \cdot {\bf R})^2  + \frac{5}{2}({\bf e}_z \cdot\blg{p})^2\vphantom{\frac{1}{2}} + 14 q ({\bf e}_z \cdot {\bf R})\,({\bf e}_z \cdot\blg{p})+ \frac{3}{2}q {\bf e}_z \cdot {\sf Q}\,{\bf e}_z\right ].
\end{align}

The motion of the particle induces the electric current $I = \dot{Q}$ in the circuit connecting the two pickup electrodes.
The current is related to the voltage $U_z$ by Kirchhoff's circuit equations, coupling the circuit dynamics to the nanoparticle motion through Eqs.~\eqref{main_force}, \eqref{main_torque}. Next, we will derive the combined circuit-nanoparticle dynamics for series and parallel RLC circuits.

\section{Resistive cooling}\label{main_resistive_cooling}

The motional energy of a charged particle can be reduced by coupling it to a series or parallel RLC circuit  through dissipation in the resistor \tcr{\cite{itano1995, dehmelt1968bolometric,church1969radiative,wineland1975principles, brown1986,major2005charged,kaltenbacher2011resistive,di2015toward,cornell1989single}}. The present section demonstrates how this resistive cooling can dampen the combined rotational-translational state of a nanoparticle and provides the resulting damping rates in the quasi-adiabatic and on-resonance limits. \tcr{Thus, it expands the theory of resistive cooling of ions to particles with rotational degrees of freedom.}

\subsection{Adiabatic cooling with series RLC circuits}\label{sec:seriescooling}

Connecting the two pick-up electrodes via a resistor and an inductor couples the nanoparticle to a series RLC circuit, see Fig.~\ref{figure1}. The inductance $L$ relates the permeating magnetic flux $\Phi$ to the current $I=\Phi/L$ so that the voltage $U_L=\dot\Phi$ drops across the inductor. The resistor $R$ at temperature $T$ induces the voltage drop $U_R=RI$ as well as fluctuations $U_{\rm fl}$ due to Johnson-Nyquist noise, with $\braket{U_{\rm fl}(t)}=0$ and $ \braket{U_{\rm fl}(t)U_{\rm fl}(t+\tau)}=2k_{\rm B}TR\delta(\tau)$. \tcr{The achievable particle temperature is thus ultimately limited by the circuit temperature.} Combining Kirchhoff's circuit law, $U_z+U_L+U_R-U_{\rm fl}=0$, with the induced charge (\ref{sec3_coupling}) yields
\begin{equation}
    \dot{\Phi}=-\frac{Q}{C}-\frac{1}{C}Q_{\rm ind}(\bls{R},\Omega)-\frac{R}{L}\Phi+U_{\rm fl}(t).
\end{equation}
This equation, together with $\dot{Q}=\Phi/L$ and Eqs.~\eqref{main_force}, \eqref{main_torque}, 
fully determines the combined nanoparticle-circuit dynamics. They are described by the Hamiltonian
\begin{align}\label{sec3_hamiltonian}
H=& H_{\rm np}+\frac{\Phi^2}{2L}+\frac{1}{2C}\left[Q+Q_{\rm ind}(\bls{R},\Omega)\right]^2,
\end{align}
together with damping and noise of the flux $\Phi$ due to the resistor,
\begin{equation}\label{main_series_circuit_dgls}
\dot Q=\frac{\partial H}{\partial \Phi},\qquad \dot \Phi=-\frac{\partial H}{\partial Q}-\frac{R}{L}\Phi+U_{\rm fl}(t).
\end{equation}

In the absence of Johnson-Nyquist noise, the charge dynamics \eqref{main_series_circuit_dgls} is described by a driven and damped harmonic oscillator,
\begin{equation}\label{app_rate_circuit_dgl}
\ddot Q+\gamma_{\rm s} \dot Q+\omega_{\rm LC}^2Q=-\omega_{\rm LC}^2Q_{\rm ind}(\bls{R},\Omega),
\end{equation}
with resonance frequency $\omega_{\rm LC}=1/\sqrt{LC}$ and damping rate $\gamma_{\rm s}=R/L$. For times much longer than the circuit relaxation time, the charge at time $t$ depends only on the particle trajectory,
\begin{align}\label{app_rate_part_sol}
Q(t)\simeq & -\frac{\omega_{\rm LC}^2}{\Delta_{\rm s}}\int_0^{\infty}d\tau\,\sin\left(\Delta_{\rm s}\tau\right)e^{-\gamma_{\rm s} \tau/2} Q_{\rm ind}[\bls{R}(t-\tau),\Omega(t-\tau)],
\end{align}
with $\Delta_{\rm s}^2 = \omega_{\rm LC}^2 - \gamma_{\rm s}^2/4$.

If the particle moves much slower than the circuit dynamics, the trajectory in \eqref{app_rate_part_sol} can be expanded to first order
\begin{align}\label{eq:adappr}
Q_{\rm ind}[\bls{R}(t-\tau),\Omega(t-\tau)]\simeq & Q_{\rm ind}[\bls{R}(t),\Omega(t)]-\tau \frac{d}{dt}Q_{\rm ind}[\bls{R}(t),\Omega(t)].
\end{align}
The circuit then follows the particle trajectory quasi-adiabatically,
\begin{align}\label{app_rate_derivatives}
Q\simeq-Q_{\rm ind}+\frac{\tcr{\gamma_s}}{\omega_{\rm LC}^2}\left[\frac{\bls{P}}{m}\cdot\frac{\partial}{\partial \bls{R}}Q_{\rm ind}+({\rm I}^{-1}\bls{J})\cdot\bls{T}\right],
\end{align}
where we used $\dot{\bf R} = {\bf P}/m$ and $\dot{{\sf R}} = ({\rm I}^{-1} {\bf J})\times {\sf R}$. Comparison of the charge \eqref{app_rate_derivatives} with the induced charge (\ref{sec3_coupling}) yields the induced voltage
\begin{equation}\label{app_QU}
U_z\simeq \frac{\tcr{\gamma_s}}{C\omega_{\rm LC}^2}\left[\frac{\bls{P}}{m}\cdot\frac{\partial}{\partial \bls{R}}Q_{\rm ind}+({\rm I}^{-1}\bls{J})\cdot\bls{T}\right].
\end{equation}
Inserting this voltage drop into the particle equations of motion \eqref{main_force}, \eqref{main_torque} thus yields a damping force and torque.

The strength of the combined rotational-translational damping can be quantified by calculating the contraction rate of an initial phase-space volume \cite{ezra2004statistical,schafer2020}. This  volume would remain constant under the dynamics described by Hamiltonian \eqref{eq:npham}, while dissipation reduces it. Since the friction force is linear in the canonical momentum coordinates, the resulting contraction rate is given by the divergence of the non-conservative part of the force and torque appearing in \eqref{main_force} and \eqref{main_torque} \cite{ezra2004statistical,rudolph2021}
\begin{align}\label{app_contractionrate1}
\Gamma_{\rm ps}=&\frac{\partial Q_{\rm ind}}{\partial \bls{R}}\cdot\frac{\partial U_z}{\partial \bls{P}}
+\sum_{\mu}\frac{\partial Q_{\rm ind}}{\partial \mu}\frac{\partial U_z}{\partial p_\mu},
\end{align}
where $\mu = \alpha,\beta,\gamma$ are the Euler angles and $p_\mu$  the corresponding canonical angular momenta, see \ref{appendix_hamiltonian}. One finds from \eqref{app_QU} that
\begin{equation}\label{main_cooling_rate}
\Gamma_{\rm ps}(\bls{R},\Omega)=R \left [ \frac{1}{m}\left(\frac{\partial Q_{\rm ind}}{\partial \bls{R}}\right)^2+\bls{T}\cdot{\rm I}^{-1}\bls{T} \right ].
\end{equation}
Here we used that the induced charge \eqref{eq:indcharge} and the torque \eqref{eq:unittorque} are related by $\partial Q_{\rm ind}/\partial \alpha=\bls{e}_z\cdot\bls{T}$, $\partial Q_{\rm ind}/\partial \beta=\bls{e}_\xi\cdot\bls{T}$, and $ \partial Q_{\rm ind}/\partial \gamma=\bls{N}_3\cdot\bls{T}$.

The contraction rate (\ref{main_cooling_rate}) is non-negative so that the circuit cools the particle motion whenever the circuit-induced force and torque are non-zero (as long as Johnson-Nyquist noise can be neglected). The efficiency of this cooling process is determined by the electrode configuration, defining the position- and orientation dependence of Eq.~\eqref{main_cooling_rate}. The optimal electrode arrangement depends on the particle charge distribution, as will be illustrated for linear and quadrupole pick-up configurations below.

\subsection{Adiabatic cooling with parallel RLC circuits}

In a parallel RLC circuit, formed by connecting the resistor and inductor in parallel to the pick-up electrodes, Johnson-Nyquist noise appears as a fluctuating current source $I_{\rm fl}(t)=U_{\rm fl}(t)/R$ in parallel to the resistor \cite{pfeifer1959elektronisches}, see Fig.~\ref{figure1}. Kirchhoff's law $I=I_L+I_R+I_{\rm fl}(t)$ with $I_L=\Phi/L$ then yields the circuit dynamics
\begin{align}\label{sec2_capacitor_charge_decay}
\dot Q&=\frac{\partial H}{\partial \Phi}-\frac{1}{RC}\left[Q+Q_{\rm ind}(\bls{R},\Omega)\right]+I_{\rm fl}(t),
\end{align}
and
\begin{equation}
\dot\Phi=-\frac{\partial H}{\partial Q},
\end{equation}
where we used that $\dot{\Phi}=-U_z$. Thus, compared to the series RLC circuit, the resistor dissipates the capacitor charge $CU_z$, rather than the flux $\Phi$. Likewise, Johnson-Nyquist noise acts on the charge rather than the current. (A canonical transformation from the charge $Q$ to the capacitor charge $Q' = C U_z$ is discussed in \ref{appendix_hamiltonian}.)

The circuit dynamics in the absence of noise are described by a damped harmonic oscillator,
\begin{align}
    \ddot Q + \gamma_{\rm p} \dot Q + \omega_{\rm LC}^2 Q =  & - \omega_{\rm LC}^2 \left [\vphantom{\frac{\gamma_{\rm p}}{\omega_{\rm LC}^2}} Q_{\rm ind}(\bls{R},\Omega)+ \frac{\gamma_{\rm p}}{\omega_{\rm LC}^2} \frac{d}{dt}Q_{\rm ind}(\bls{R},\Omega) \right ],
\end{align}
with damping rate $\gamma_{\rm p} = 1/RC$. In contrast to the series RLC circuit \eqref{app_rate_circuit_dgl}, the charge in the parallel RLC circuit is driven not only by the  position and orientation of the particle but also by its linear and angular velocities. This yields the solution
\begin{align}
Q(t) \simeq & -\frac{\omega_{\rm LC}^2}{\Delta_{\rm p}}\int_0^{\infty}d\tau\,\sin\left(\Delta_{\rm p}\tau\right)e^{-\gamma_{\rm p} \tau/2}\left[Q_{\rm ind}[\bls{R}(t-\tau),\Omega(t-\tau)] \vphantom{\frac{\gamma_{\rm p}}{\omega_{\rm LC}^2}}\right. \nonumber\\
& \left. + \frac{\gamma_{\rm p}}{\omega_{\rm LC}^2} \frac{d}{dt}Q_{\rm ind}[\bls{R}(t-\tau),\Omega(t-\tau)]\right].
\end{align}
Approximating the particle motion by its instantaneous linear  and angular velocity \eqref{eq:adappr} one finds that the charge offset turns independent of the velocities,
\begin{align}
    Q(t) \simeq & - Q_{\rm ind}[\bls{R}(t),\Omega(t)].
\end{align}
The voltage offset and the nanoparticle damping rate thus vanish in the quasi-adiabatic limit, $\Gamma_{\rm ps} \approx 0$.
This is because the current flows through the coil rather than through the resistor as in the series RLC circuit. As shown below, a parallel RLC circuit can effectively cool a harmonically trapped nanoparticle if the circuit frequency is on resonance with the motion of the particle.

\subsection{Linear pick-up configuration}\label{sec:lc}

A linear pick-up configuration is realized if the two electrodes   are placed at opposite sides of the trap, with a distance of $2z_0$, see Fig.~\ref{figure1}(a). This can for instance be implemented by connecting the endcap electrodes of a Paul trap (see below). The resulting reference potential is approximately linear in the trapping region,
\begin{equation}
    \Phi_0(\bls{r})=\frac{k_1}{z_0}\bls{e}_z\cdot\bls{r}
\end{equation}
so that the the charge induced by the nanoparticle is determined by its monopole and dipole moment,
\begin{equation}
    Q_{\rm ind}(\bls{R},\Omega)=\frac{k_1}{z_0}\bls{e}_z\cdot(q\bls{R}+\blg{p}).
\end{equation}
Here we defined $z=0$ as the center of the electrode arrangement and chose the coordinate system such that $-\bls{e}_z$ is the direction of the electrode-induced electric field in the trapping region. The numerical factor $k_1$ accounts for the shape of the electrodes;  an infinitely extended plate capacitor yields $k_1=1/2$. Note that the presence of further electrodes would modify both $k_1$ and, in asymmetric setups, the voltage offset.

The force and torque on the nanoparticle are given by Eqs.~\eqref{main_force}, \eqref{main_torque} with
\begin{equation}
    \bls{T}(\Omega)=\frac{k_1}{z_0}\blg{p}\times\bls{e}_z.
\end{equation}
In a series RLC circuit, this gives rise to the adiabatic damping rate
\begin{equation}\label{eq:ldr}
\Gamma_{\rm ps}(\Omega)=\frac{Rk_1^2}{z_0^2}\left[\frac{q^2}{m}+(\blg{p}\times \bls{e}_z)\cdot{\rm I}^{-1}(\blg{p}\times \bls{e}_z)\right],
\end{equation}
where the first and second term is due to cooling of the center of mass \tcr{\cite{itano1995,kaltenbacher2011resistive,di2015toward,goldwater2018levitated}} and the rotations, respectively. Note that rotational cooling vanishes whenever the dipole vector is aligned with the electric field generated by the circuit, and that the linear pick-up configuration cannot cool rotations around the dipole axis of the particle.

The cooling rate \eqref{eq:ldr} is independent of the particle position. This is in contrast to the quadrupole pick-up configuration discussed below and to optical cavity cooling schemes \cite{chang2010,romeroisart2010,barker2010,stickler2016a}, where an optical tweezer ensures that the particle does not enter regions of vanishing cooling rate. The rate is proportional to the square of the total charge, rendering resistive cooling attractive for massive and highly charged particles \cite{goldwater2018levitated}. For instance a $10^6$\,u particle with a realistic loading of $120$ elementary charges \cite{draine1987} can be cooled as fast as a ${^{88}\rm Sr}^+$ ion.

\subsection{Quadrupole pick-up configuration}\label{sec:qc}

A quadrupole pick-up configuration consists of two pairs of electrodes placed around the trapping region, so that opposite electrodes are at the same potential and neighboring electrodes are connected via the circuit, see Fig.~\ref{figure1}(b). This can be implemented by connecting the rods in a linear Paul trap via an RLC circuit to the endcap electrodes. This configuration is associated with a quadrupole field in the trapping region,
\begin{equation}
    \Phi_0(\bls{r})=\frac{k_2}{2z_0^2}\bls{r}\cdot\mathsf{G}\,\bls{r},
\end{equation}
characterized by the lengthscale $z_0$  and a traceless geometry tensor with real eigenvalues $g_i$ and orthogonal eigenvectors ${\bf g}_i$,
\begin{equation}\label{eq:gtensor}
    \mathsf{G}=\sum_{i=1}^3 g_i \bls{g}_i\otimes\bls{g}_i.
\end{equation}
We choose $k_2$ such that the maximum absolute value of the $g_i$ is on the order of unity.

The charge induced by the nanoparticle follows from Eq.~\eqref{eq:indcharge} as
\begin{align}
Q_{\rm ind}(\bls{R},\Omega)= & \frac{k_2}{2z_0^2}\left(q\bls{R}\cdot \mathsf{G}\,\bls{R}+2\blg{p}\cdot\mathsf{G}\,\bls{R} \vphantom{\frac{1}{3}\sum_{i=1}^3}+\frac{1}{3}\sum_{i=1}^3 g_i\bls{g}_i\cdot\mathsf{Q}\, \bls{g}_i\right).
\end{align}
This implies a torque per unit voltage of
\begin{equation}
    \bls{T}(\bls{R},\Omega)=\frac{k_2}{z_0^2}\left(\blg{p}\times\mathsf{G}\,\bls{R}-\frac{1}{3}\sum_{i=1}^3g_i\bls{g}_i\times\mathsf{Q}\,\bls{g}_i\right)
\end{equation}
and the adiabatic damping rate
\begin{align}
    \Gamma_{\rm ps}(\bls{R},\Omega)=&\frac{Rk_2^2}{z_0^4}\left[\frac{1}{m}(q\bls{R}+\blg{p})\cdot\mathsf{G}^2(q\bls{R}+\blg{p})\vphantom{\sum_{i=1}^3\frac{g_i}{3}}+\left(\blg{p}\times\mathsf{G}\bls{R}-\sum_{i=1}^3\frac{g_i}{3}\bls{g}_i\times\mathsf{Q}\bls{g}_i\right)\right.\nonumber \\
    &\left.\cdot {\rm I}^{-1}\left(\blg{p}\times\mathsf{G}\bls{R}-\sum_{i=1}^3\frac{g_i}{3}\bls{g}_i\times\mathsf{Q}\bls{g}_i\right)\right].
\end{align}

As in the linear configuration, the center-of-mass damping rate is proportional to $q^2/M$, but it here vanishes in the center of the electrode arrangement, at ${\bf R} = 0$. Rotational cooling is determined by the particle dipole and quadrupole orientation. Importantly, asymmetric quadrupole pick-up configurations ($g_1 \neq g_2\neq g_3$) enable rotational cooling of all orientational degrees of freedom even for vanishing dipole moments provided the quadrupole tensor has three distinct eigenvalues.

\subsection{Cooling of harmonically trapped rotors}\label{sec_frequency_dependent_cooling_rates}

For sufficiently small amplitudes of the rotational and centre-of-mass motion in comparison, the normal modes are coupled linearly to the circuit. Solving the circuit equations of motion in Fourier space then yields frequency-dependent damping rates for each normal coordinate. In the following, we first illustrate this for a single center-of-mass mode in the linear pick-up configuration \tcr{\cite{itano1995,kaltenbacher2011resistive,goldwater2018levitated}}, before discussing the generalization to more than one mechanical degree of freedom\tcr{, including libration,} and arbitrary electrode arrangements \tcr{with arbitrary circuit impedance}.

For simplicity, we take  the nanoparticle to move only in the  direction separating the pick-up electrodes, and assume the trapping and image forces to give rise to a harmonic potential of frequency $\omega_0$. In Fourier space, the equation of motion then reads
\begin{equation}\label{eq:fteom}
    - m \omega^2 \widetilde{z}(\omega) + m \omega_0^2 \widetilde{z}(\omega) = -\frac{q k_1}{z_0} \widetilde{U}_z(\omega),
\end{equation}
where $\widetilde{z}(\omega)$ denotes the Fourier transform of the particle position $z(t)$. The  circuit impedance $Z(\omega)$ relates the voltage offset $\widetilde{U}_z(\omega) = -Z(\omega) \widetilde{I}(\omega)$  to the induced current. 

Since we neglect the rotational motion, the induced voltage can be determined from Eq.~\eqref{sec3_coupling} as
\begin{equation}\label{sec_3E_Uz}
    \widetilde{U}_z(\omega) = i \omega \frac{q k_1}{z_0}\frac{\widetilde{z}(\omega) Z(\omega)}{1 + i \omega C Z(\omega)}.
\end{equation}
Inserting this into the equation of motion \eqref{eq:fteom} and using that the oscillator moves harmonically with frequency $\omega_0$ gives the damping rate
\begin{equation}\label{sec3_dampingrate_arbitrary}
\Gamma(\omega_0)\simeq\frac{q^2k_1^2}{mz_0^2}{\rm Re}\!\left[\frac{Z(\omega_0)}{1+iC \omega_0  Z(\omega_0)}\right].
\end{equation}

For instance, for a series RLC circuit with $Z(\omega)=R+i\omega L$ the damping rate is
\begin{subequations}\label{sec3B_cooling_rates}
\begin{equation}\label{cooling_rate_series}
\Gamma_{\rm s}(\omega_0)=\frac{q^2k_1^2}{Cmz_0^2}\frac{\gamma_{\rm s}\omega_{\rm LC}^2}{\omega_0^2\gamma_{\rm s}^2+(\omega_0^2-\omega_{\rm LC}^2)^2}.
\end{equation}
It takes its maximum value at $\omega_0 = 0$ if $\gamma_{\rm s}$ is greater than $ \sqrt{2}\omega_{\rm LC}$ and else at $\omega_0^2 = \omega_{\rm LC}^2 - \gamma_{\rm s}^2/2$. The value of $\Gamma_{\rm s}$ at $\omega_0 = 0$ coincides with the quasi-adiabatic damping rate \eqref{main_cooling_rate}.

For the parallel RLC circuit with $1/Z(\omega)=1/R-i/\omega L$ one obtains
\begin{equation}\label{cooling_rate_parallel}
\Gamma_{\rm p}(\omega_0)=\frac{q^2k_1^2}{Cmz_0^2}\frac{\gamma_{\rm p}\omega_0^2}{\omega_0^2\gamma_{\rm p}^2+(\omega_0^2-\omega_{\rm LC}^2)^2}.
\end{equation}
\end{subequations}
On resonance, where $\omega_0 = \omega_{\rm LC}$, the rate attains the maximum value $Rq^2k_1^2/mz_0^2$ \tcr{\cite{itano1995,kaltenbacher2011resistive,di2015toward,goldwater2018levitated}}, which coincides with the quasi-adiabatic damping rate in the series RLC circuit at zero frequency. In contrast, the damping rate in the parallel RLC circuit goes to zero for $\omega_0 = 0$, as in the adiabatic case.

In the case of several harmonically trapped degrees of freedom, which are linearly coupled to the circuit, one obtains damping rates of the form \eqref{sec3_dampingrate_arbitrary} for each of them. In addition, the circuit mediates linear coupling between the normal modes of the trapping potential. If this coupling leaves the normal mode frequencies sufficiently distinct, tuning the circuit can be used to selectively cool single degrees of freedom.

A straightforward generalization of \eqref{sec3_dampingrate_arbitrary} for an arbitrary electrode geometry yields for small center-of-mass oscillations
\begin{subequations}\label{freqfrictiontensors}
\begin{equation}
\mathsf{\Gamma}_{\rm cm}(\omega)\simeq{\rm Re}\!\left[\frac{Z(\omega)}{1+iC \omega  Z(\omega)}\right]\frac{1}{m}\left (\frac{\partial Q_{\rm ind}}{\partial \bls{R}}\otimes \frac{\partial Q_{\rm ind}}{\partial \bls{R}}\right ).
\end{equation}
or for small librations
\begin{equation}
\mathsf{\Gamma}_{\rm rot}(\omega)\simeq{\rm Re}\!\left[\frac{Z(\omega)}{1+iC \omega Z(\omega)}\right](\bls{T}\otimes \bls{T}){\rm I}^{-1}.
\end{equation}
\end{subequations}
Here we assumed that translational and rotational modes do not hybridize in the trapping potential. All tensors are evaluated at the particle equilibrium position and orientation. Note that the rotational friction tensor is not symmetric in general, which is also the case in a gaseous environment \cite{martinetz2018}. 

Momentum and angular-momentum diffusion naturally arise if the fluctuating voltages and currents in the circuit equations of motion are taken into account. According to the fluctuation-dissipation theorem \cite{gardiner1985}, the voltage fluctuations $\widetilde{U}_{z,{\rm fl}}(\omega)$ between the capacitor plates are determined by the total impedance $Z(\omega)/[1 + i \omega C Z(\omega)]$, so that $\langle \widetilde{U}_{z,{\rm fl}}(\omega_0)\rangle=0$ and
\begin{equation}
    \langle \widetilde{U}_{z,{\rm fl}}(\omega)\widetilde{U}^*_{z,{\rm fl}}(\omega')\rangle=\frac{k_{\rm B}T}{\pi}{\rm Re}\!\left[\frac{Z(\omega)}{1+iC \omega  Z(\omega)}\right]\delta(\omega-\omega')\tcr{,}
\end{equation}
\tcr{characterized by the circuit temperature $T$.} Adding the fluctuating voltage to Eq.~(\ref{sec_3E_Uz}) \tcr{and repeating the steps above leading to Eqs.~(\ref{freqfrictiontensors})} yields the effective momentum diffusion tensors
\begin{align}
\mathsf{D}_{\rm cm}(\omega)=&k_{\rm B}Tm\mathsf{\Gamma}_{\rm cm}(\omega),\\
\mathsf{D}_{\rm rot}(\omega)=&k_{\rm B}T\mathsf{\Gamma}_{\rm rot}(\omega){\rm I}.
\end{align}
\tcr{The noise is thus effectively white and proportional to $T$. This implies that the degrees of freedom coupled to the circuit will ultimately thermalize to the circuit temperature on timescales quantified by the friction tensors (\ref{freqfrictiontensors}).}

\tcr{To obtain a realistic description of nanorotor cooling in Paul traps, one must include the influence of the time-dependent trapping fields on the particle motion (see Sec.~\ref{sec_electric_levitation}), and account for other noise sources, such as gas collisions (see Secs.~\ref{sec:equilibration} and \ref{sec:exp}).} 

\section{Paul trap dynamics}\label{sec_electric_levitation}

The trapping potential $V_{\rm tr}$, that is required to levitate the nanoparticle between the pickup electrodes, has been left unspecified so far. While optical or magnetic fields can be used for that purpose, the most common means of levitating a charged particle are the alternating electric fields of a Paul trap. In this section, we therefore derive the coupled translational and rotational macromotion in a general quadrupole ion trap, and identify the time-independent, effective potential. This will be used in Sec.~\ref{sec:exp} to simulate  the cooling dynamics, in order to demonstrate that Paul traps offer a viable platform for implementing the resistive cooling and circuit control of nanorotors.

\subsection{Time-dependent force and torque}
The electric quadrupole field at the center of a Paul trap has the general form \cite{leibfried2003quantum},
\begin{equation}\label{sec_2_electric_pot}
{\bf E}(\bls{R},t)= - \frac{U(t)}{\ell_0^2}\mathsf{A}\,\bls{R},
\end{equation}
where $U(t)=U_{\rm dc}+U_{\rm ac}\cos(\omega_{\rm ac} t)$ is the applied voltage with drive frequency $\omega_{\rm ac}$. The arrangement of the electrodes is characterized by the   real, symmetric, and traceless tensor
    \begin{equation}\label{eq:Atensor} 
    \mathsf{A}=\sum_{i = 1}^3 a_i\bls{a}_i\otimes\bls{a}_i,
\end{equation}
and by the length scale $\ell_0$ of the trapping field, chosen such that the maximum absolute value of the eigenvalues $a_i$ is on the order of unity. (The form of $\mathsf{A}$ for special Paul trap geometries can be found in Sect.~\ref{eff_pot_special_particle}.)

\begin{figure*}
\centering
\includegraphics[width = 1\textwidth]{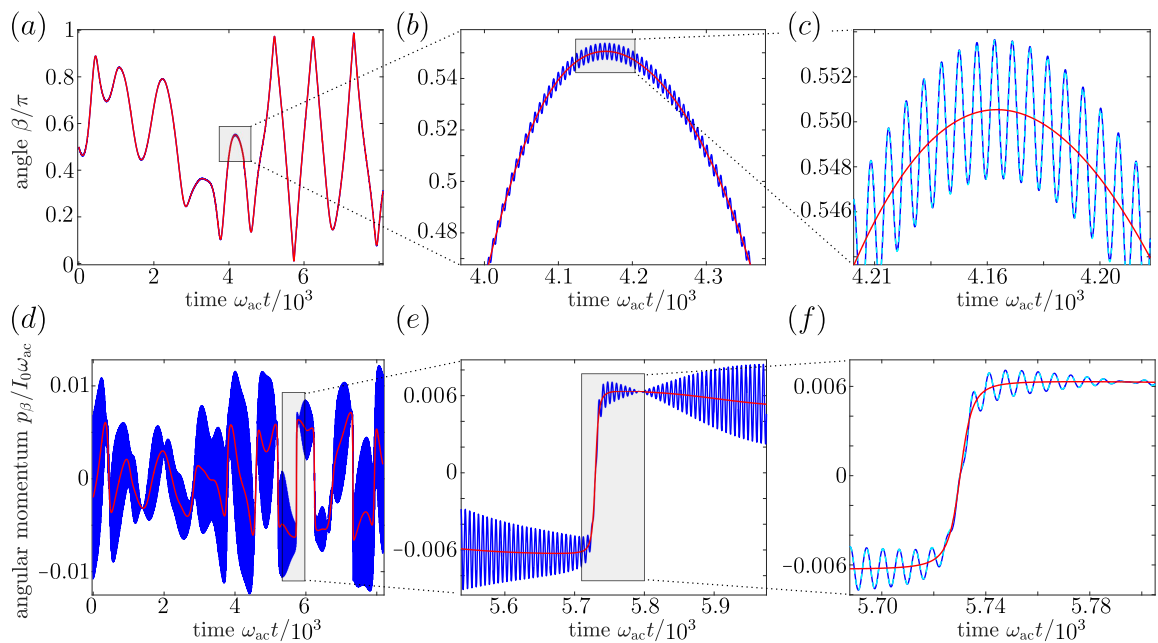}
\caption{Dynamics of a charged particle in the time-dependent quadrupole field (\ref{sec_2_electric_pot}) of a ring-shaped Paul trap. (a) The rotational trajectory of the polar angle $\beta$ shows very good agreement between the  exact (dark blue) and the effective (red) dynamics due to the effective potential (\ref{sec_2_effective_potential}) for high frequencies $\omega_{\rm ac}$. (b)-(c) The small oscillations around the macromotion trajectory are well described by \eqref{sec2_delta} (light blue dashed line in (c)); they vanish for increasing $\omega_{\rm ac}$. (d)-(f) The exact dynamics of the canonically conjugate momentum $p_\beta$ (dark blue) deviate significantly from the macromotion (red), but they are in very good agreement when the micromotion is included \eqref{korrektur_mom_und_ang_mom} (light blue dashed line in (f)).  The simulation details and a centre-of-mass trajectory are provided in \ref{app_simulation}.} \label{figure2}
\end{figure*}

The trapping force and torque due to the time-dependent electric field \eqref{sec_2_electric_pot} are given by
\begin{subequations}\label{sec2_force_and_torque}
\begin{align}\label{sec2_force}
\bls{F}_{\rm tr}&=-\frac{U(t)}{\ell_0^2}(q\mathsf{A}\bls{R}+\mathsf{A}\blg{p})
\\
\label{sec2_torque}
{\bf N}_{\rm tr}&=-\frac{U(t)}{\ell_0^2}\left[\blg{p}\times\mathsf{A}\bls{R}-\frac{1}{3}\sum_{i = 1}^3a_i\bls{a}_i\times\mathsf{Q}\bls{a}_i\right].
\end{align}
\end{subequations}

In  general, the induced ro-translational dynamics will be rather complicated and strongly coupled for non-vanishing electric dipole moments. However, we will see in the following  that for sufficiently large driving frequencies, and if the AC voltage dominates, the nanoparticle center of mass is stably trapped. The ro-translational motion then decomposes into a slowly varying, large amplitude macromotion $\bls{r}$ and $\bls{n}_i$ of the center-of-mass position and the orientation, and a small amplitude, rapidly oscillating micromotion, $\blg{\epsilon}$ and $\blg{\delta}$.

\subsection{Micromotion}

Inserting the macro-micro separation ansatz $\bls{R}=\bls{r}+\boldsymbol{\epsilon}$ and $\bls{N}_i=\bls{n}_i+\boldsymbol{\delta}\times \bls{n}_i$ into Eq.~(\ref{sec2_force}) and neglecting the small quantities $|\blg{\epsilon}|\ll |\bls{r}|$ and $|\blg{\delta}|\ll 1$ one obtains from Newton's equation for the center of mass
\begin{align}\label{sec2_force_short_timescale}
m\ddot{\blg{\epsilon}}+m\ddot{\bls{r}}\simeq&-\frac{U_{\rm dc}}{\ell_0^2}\left(q\mathsf{A}\bls{r}+\sum_{i = 1}^3 p_i\mathsf{A}\bls{n}_i\right)-\cos(\omega_{\rm ac} t)\frac{U_{\rm ac}}{\ell_0^2}\left(q\mathsf{A}\bls{r}+\sum_ip_i\mathsf{A}\bls{n}_i\right).
\end{align}
The second term on the left-hand side is negligibly small due to the assumed separation into a macro- and micromotion. On the right hand-side, the first term, which describes the DC force acting at the macromotion position,  is dominated by the AC contribution in the second term. Integrating the remaining equation and using that the macromotion coordinates do not change on the timescale of the micromotion gives the zero-mean center-of-mass micromotion $\blg{\epsilon}\simeq \blg{\epsilon}_0\cos(\omega_{\rm ac} t)$ with
\begin{equation}\label{sec2_epsilon}
\blg{\epsilon}_0=\frac{U_{\rm ac}}{m\omega_{\rm ac}^2\ell_0^2}\left(q\mathsf{A}\bls{r}+\sum_ip_i\mathsf{A}\bls{n}_i\right).
\end{equation}

In order to calculate rotational micromotion $\blg{\delta}$, we insert the separation ansatz into Eq.~(\ref{sec2_torque}) and use the kinematic relation
\begin{align}\label{sec2_angular_momentum_change}
\dot{\bls{J}}= & \frac{1}{2}\sum_{i = 1}^3{\rm I}(\bls{N}_i\times\ddot{\bls{N}}_i)+\frac{1}{4}\sum_{i,j = 1}^3(\bls{N}_i\times\dot{\bls{N}}_i)\times{\rm I}(\bls{N}_j\times\dot{\bls{N}}_j),
\end{align}
 in Newton's equation for the angular momentum. Neglecting all small terms under the same assumptions as above, gives the approximate equation for the rotational micromotion
\begin{align}
\sum_{i = 1}^3 I_i\bls{n}_i(\bls{n}_i\cdot\ddot{\blg{\delta}})\simeq & -\frac{U_{\rm ac}\cos(\omega_{\rm ac} t)}{\ell_0^2}\left[\sum_{i = 1}^3 p_i\bls{n}_i\times\mathsf{A}\bls{r}\vphantom{\frac{1}{3}\sum_{i,\ell, m = 1}^3}-\frac{1}{3}\sum_{i,\ell,m = 1}^{3}a_iQ_{\ell m}\bls{a}_i\times\bls{n}_\ell(\bls{n}_m\cdot\bls{a}_i)\right].
\end{align}
This yields the rapidly oscillating, zero-mean rotational micromotion $\blg{\delta}\simeq \blg{\delta}_0\cos(\omega_{\rm ac} t)$ with amplitude
\begin{align}\label{sec2_delta}
\blg{\delta}_0=\frac{U_{\rm ac}}{\omega_{\rm ac}^2\ell_0^2}\sum_{j = 1}^3\frac{1}{I_j}\bls{n}_j \left[\bls{n}_j\cdot\left(\sum_{i = 1}^3p_i\bls{n}_i\times\mathsf{A}\bls{r}-\frac{1}{3}\sum_{i,\ell, m = 1}^3 a_iQ_{\ell m}\bls{a}_i\times\bls{n}_\ell(\bls{n}_m\cdot\bls{a}_i)\right)\right]
\end{align}

According to Eq.\ (\ref{sec2_epsilon})  the assumption $|\blg{\epsilon}|\ll |\bls{r}|$ is fulfilled provided the centre-of-mass Mathieu parameter for the charge and the dipole moment  are small, i.e. for
\begin{align}
    \frac{U_{\rm ac}q}{m\omega_{\rm ac}^2\ell_0^2} &\ll 1,
    &
    \frac{U_{\rm ac}|\blg{p}|}{m\omega_{\rm ac}^2\ell_0^2\ell_{\rm cm}} &\ll 1,
\end{align}
where $\ell_{\rm cm}$ is the length scale  of the center-of-mass motion. Similarly, Eq. (\ref{sec2_delta}) shows that the requirement $|\blg{\delta}|\ll 1$ is controlled by the rotational analogues of the Mathieu parameters
\begin{align}
    \frac{U_{\rm ac} |Q_{\ell m}|}{ I_j \omega_{\rm ac}^2 \ell_0^2} &\ll 1,
    &
    \frac{U_{\rm ac} |\blg{p}|\ell_{\rm cm}}{ I_j \omega_{\rm ac}^2 \ell_0^2} &\ll 1.
\end{align}

\subsection{Macromotion}

The center-of-mass macromotion force follows from inserting the micromotion trajectories $\blg{\epsilon}$ and $\blg{\delta}$ into the force (\ref{sec2_force}) and time-averaging the equation of motion over one field oscillation. Neglecting small terms and identifying ${\bf F}_{\rm eff}=m \ddot{\bf r}$ one obtains the effective force,
\begin{equation}\label{sec2_eff_force}
{\bf F}_{\rm eff}\simeq-\frac{U_{\rm dc}}{\ell_0^2}\mathsf{A}(q\bls{r}+\blg{p})-\frac{U_{\rm ac}}{2\ell_0^2}\mathsf{A}\left(q\blg{\epsilon}_0+\blg{\delta}_0\times\blg{p}\right).
\end{equation}
For notational simplicity, we here re-defined $\blg{p}$, $\mathsf{Q}$ and ${\rm I}$ as the dipole moment, quadrupole tensor and inertia tensor of the macromotion, e.g. $\blg{p}=\sum_ip_i\bls{n}_i$. The second term on the right-hand side of Eq.~(\ref{sec2_eff_force}) describes the effective force due to AC driving of the trap electrodes; the first term describes the DC force.

Likewise, we insert the micromotion into the torque (\ref{sec2_torque})  and average the equation of motion over one Paul trap cycle. In order to identify the effective torque, we use the macromotion version of the kinematic relation \eqref{sec2_angular_momentum_change}.
This yields
\begin{align}\label{sec_2_effective_torque}
{\bf N}_{\rm eff} \simeq &-\frac{U_{\rm dc}}{\ell_0^2}\left[\blg{p}\times\mathsf{A}\bls{r}-\frac{1}{3}\sum_{i=1}^3a_i\bls{a}_i\times\mathsf{Q}\bls{a}_i\right]-\frac{U_{\rm ac}}{2\ell_0^2}\left[\vphantom{\sum_{i=1}^3}\left(\blg{\delta}_0\times \blg{p}\right)\times\mathsf{A}\bls{r}+\blg{p}\times\mathsf{A}\blg{\epsilon}_0\right.\nonumber \\
&\left.-\frac{1}{3}\sum_{i=1}^3a_i\bls{a}_i\times \mathsf{Q}\left(\bls{a}_i\times\blg{\delta}_0\right)-\frac{1}{3}\sum_{i = 1}^3a_i\bls{a}_i\times \left(\blg{\delta}_0\times\mathsf{Q}\bls{a}_i\right)\right].
\end{align}

The force \eqref{sec2_eff_force} and the torque \eqref{sec_2_effective_torque} are both described by the time-independent effective potential  (or pseudopotential)
\begin{align}\label{sec_2_effective_potential}
V_{\rm eff}({\bf r},\Omega) =&\frac{U_{\rm dc}}{\ell_0^2}\left(\frac{q}{2}\bls{r}\cdot\mathsf{A}\bls{r}+\blg{p}(\Omega)\cdot\mathsf{A}\bls{r}+\frac{1}{6}\sum_{i=1}^3a_i\bls{a}_i\cdot\mathsf{Q}(\Omega)\bls{a}_i\right)\nonumber\\
&+\frac{U_{\rm ac}^2}{4m\omega_{\rm ac}^2\ell_0^4}(q\bls{r}+\blg{p}(\Omega))\cdot\mathsf{A}^2(q\bls{r}+\blg{p}(\Omega))\nonumber\\
&+\frac{U_{\rm ac}^2}{4\omega_{\rm ac}^2\ell_0^4}\left(\blg{p}(\Omega)\times\mathsf{A}\bls{r}-\frac{1}{3}\sum_{i =1}^3a_i\bls{a}_i\times\mathsf{Q}(\Omega)\bls{a}_i\right)\nonumber\\
&\cdot{\rm I}^{-1}(\Omega)\left(\blg{p}(\Omega)\times\mathsf{A}\bls{r}-\frac{1}{3}\sum_{i=1}^3a_i\bls{a}_i\times\mathsf{Q}(\Omega)\bls{a}_i\right).
\end{align}
Here, ${\bf r}$ is the macromotion centre-of-mass coordinate and  $\Omega$ denotes from now on the macromotion orientation. One can verify the equivalence of (\ref{sec_2_effective_potential}) with (\ref{sec2_eff_force}) and (\ref{sec_2_effective_torque}) by parametrizing the translational and rotational degrees of freedom and comparing Hamilton's and Newton's equations in a long but straightforward calculation.

The effective potential (\ref{sec_2_effective_potential}) provides an accurate description of the time-averaged particle motion for high AC frequencies, replacing the complicated ro-translational dynamics in the time-dependent field (\ref{sec2_force_and_torque}). This is illustrated in Figs.~\ref{figure2} and \ref{figure6}. If the dipole and quadrupole moments vanish, Eq.~(\ref{sec_2_effective_potential}) turns into the well-known secular potential for a point charge \cite{dehmelt1968radiofrequency}, derived from a Floquet ansatz in the high-frequency limit \cite{major2005charged}. The potential (\ref{sec_2_effective_potential}) can also be derived quantum mechanically by adapting the method outlined in \cite{cook1985quantum} for the combined rotational and translational nanoparticle motion.

While the approximations $\bls{R}\simeq\bls{r}$ and $\bls{N}_i\simeq\bls{n}_i$ hold very well for sufficiently large trap frequencies $\omega_{\rm ac}$,  agreement in momentum and angular momentum is only achieved by also accounting for the micromotion, $\dot{\bls{R}}\simeq \dot{\bls{r}}-\omega_{\rm ac}\blg{\epsilon}_0\sin(\omega_{\rm ac} t)$ and $\dot{\bls{N}}_i\simeq \dot{\bls{n}}_i-\omega_{\rm ac}\blg{\delta}_0\times\bls{n}_i\sin(\omega_{\rm ac} t)$, see Fig. \ref{figure2}. The center-of-mass momentum and the angular momentum then read as
\begin{subequations}\label{korrektur_mom_und_ang_mom}
\begin{align}\label{korrektur_momentum_micromotion}
\bls{P}\simeq\;&\, m\dot{\bls{r}}-\sin(\omega_{\rm ac}t)\frac{U_{\rm ac}}{\omega_{\rm ac}\ell_0^2}\mathsf{A}\left(q\bls{r}+\blg{p}\right),
\\
\bls{J}\simeq\;&\,{\rm I} \blg{\omega}-\sin(\omega_{\rm ac}t)\frac{U_{\rm ac}}{\omega_{\rm ac}\ell_0^2}
\Big(\blg{p}\times\mathsf{A}\bls{r}-\frac{1}{3}\sum_{i = 1}^3a_i\bls{a}_i\times\mathsf{Q}\bls{a}_i\Big).
\end{align}
\end{subequations}
where $\blg{\omega}=\sum_i\bls{n}_i\times\dot{\bls{n}}_i/2$ is the macromotion angular-velocity vector. We next evaluate the effective potential (\ref{sec_2_effective_potential}) for linear and hyperbolic Paul traps, the two most commonly employed quadrupole trap geometries.

\section{Typical trap geometries}\label{eff_pot_special_particle}
The effective potential (\ref{sec_2_effective_potential}) simplifies significantly for special particle shapes and trap geometries. In the following, we consider a cylindrically symmetric particle whose body-fixed $3$-axis is the symmetry axis $\bls{m}\equiv\bls{n}_3$. The inertia tensor ${\rm I}=I(\mathbb{1}-\bls{m}\otimes\bls{m})+I_3\bls{m}\otimes\bls{m}$ involves the moment $I_3$ for rotations around the  $\bls{m}$ axis, and the moment $I$ for rotations around the perpendicular axes. The surface charge distribution is also cylindrically symmetric, so that the dipole moment is given by $\blg{p}=p_3\bls{m}$ and the quadrupole tensor by $\mathsf{Q}=Q_{3} (3\bls{m}\otimes\bls{m}-\mathbb{1})$ with $Q_3 = Q_{33}/2$. For this particle, we  now determine the effective potential \eqref{sec_2_effective_potential} in ring-shaped and linear trap geometries. Corresponding expressions for planar trap geometries \cite{alda2016trapping,hughes2011microfabricated} can be obtained in a similar fashion.

\begin{figure*}
\centering
\includegraphics[width = 0.9\textwidth]{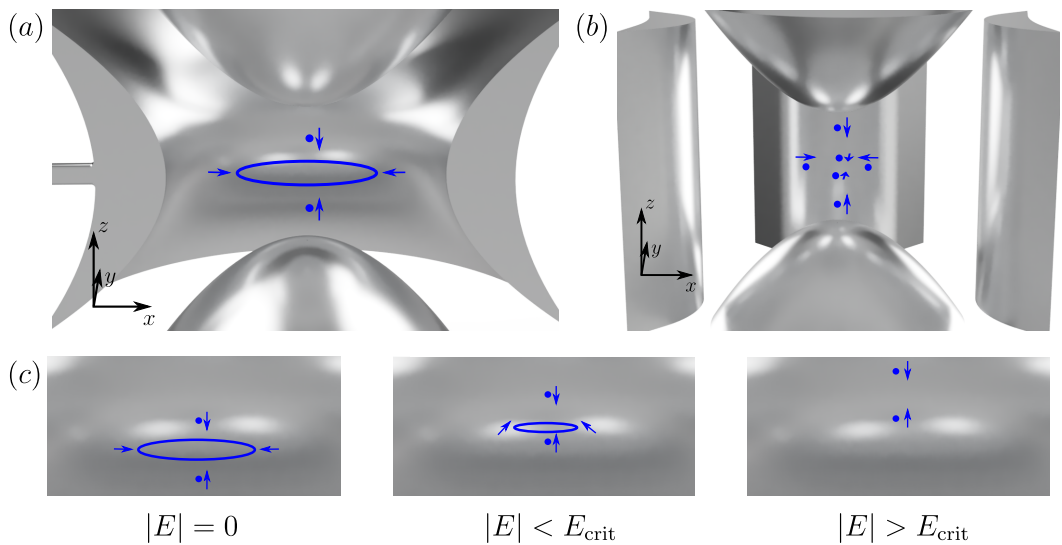}
\caption{(a) The potential minima of a cylindrically symmetric particle in a ring-shaped Paul trap are located at $\bls{r}=-\blg{p}/q$, where the dipole vector $\blg{p}$ is either aligned with the symmetry axis $\bls{e}_z$ of the Paul trap or lies in the perpendicular plane. The latter gives rise to a continuous, ring-shaped set of minima. The arrows indicate the direction of $\blg{p}/q$. (b) Also in the linear Paul trap, the trapping position and orientation of the particle are related by $\bls{r}=-\blg{p}/q$. Depending on the values of $U_{\rm ec}$, $p_3$ and $Q_3$, the dipole vector $\blg{p}$ is either aligned in the direction of $ \bls{e}_z$ or of $\pm \bls{e}_{x,y}$. (c) The presence of an additional homogeneous electric field $E{\bf e}_z$ shifts the potential minima in the ring-shaped Paul trap.  Beyond the critical field strength \eqref{eq:critvalue}, the ring minima merge with an isolated one. The distances from the trap center are exaggerated for better illustration.} \label{figure3}
\end{figure*}

\subsection{Ring-shaped Paul traps} 

Ring-shaped Paul traps are composed of a ring electrode with radius $\ell_0$ and two endcaps at distance $\ell_0/\sqrt{2}$ from the center, see Fig.~\ref{figure3}. Applying the voltage $U(t)$ between the ring and the endcaps, gives rise to an electric quadrupole field \eqref{sec_2_electric_pot} with $\mathsf{A}=\mathbb{1}-3\bls{e}_z\otimes\bls{e}_z$. Such traps are for instance used in the experiments reported in \cite{millen2015cavity,delord2017a}.

Setting the DC voltage to zero, the effective potential \eqref{sec_2_effective_potential} for the symmetric particle takes the form
\begin{align}\label{appendix_potential_ringtrap}
V_{\rm eff}({\bf r},\Omega)=&\frac{U_{\rm ac}^2}{4m\omega_{\rm ac}^2\ell_0^4}(q\bls{r}+p_3\bls{m})\cdot\mathsf{A}^2(q\bls{r}+p_3\bls{m})+\frac{U_{\rm ac}^2}{4I\omega_{\rm ac}^2\ell_0^4}\left [\bls{m}\times \mathsf{A}\left ( p_3\bls{r}+Q_3\bls{m}\right )\right]^2.
\end{align}
The potential is  positive, since $\mathsf{A}^2$ is positive definite, except for the global minima, where it vanishes. The first term in (\ref{appendix_potential_ringtrap}) is zero if the motional dipole moment compensates the permanent dipole moment, $\bls{r}=-p_3\bls{m}/q$. At this position, the second term of the potential \eqref{appendix_potential_ringtrap} vanishes for all orientations if $p_3^2=qQ_3$. For all other values of $p_3$, the particle tends to align its symmetry axis $\bls{m}$ parallel or perpendicular to the Paul-trap axis $\pm\bls{e}_z$. This is illustrated in Fig.~\ref{figure3}(a).

An additional homogeneous electric field $E{\bf e}_z$ adds the potential energy $-E\bls{e}_z\cdot(q\bls{r}+\blg{p})$ to (\ref{appendix_potential_ringtrap}), shifting the trap minima in position and orientation, see Fig. \ref{figure3}(c). At the  critical field strength
\begin{equation}\label{eq:critvalue}
E_{\rm crit}=\left | \frac{3U_{\rm ac}^2(p_3^2-qQ_3)}{mp_3\omega_{\rm ac}^2\ell_0^4}\right |,
\end{equation}
the ring minima merge with an isolated minimum, so that for $|E| > E_{\rm crit}$ only two minima remain. This implies that additional linear potentials can align the particle with the $\bls{e}_z$ axis.

\subsection{Linear Paul traps}\label{secV:linear_trap}

The linear Paul trap consists of four parallel rods of hyperbolic shape, aligned parallel to the $\bls{e}_z$-axis and placed at the corners of a square with diagonal distance $2\ell_0$, see Fig.~\ref{figure3}. Rods opposite to each other are held at equal voltage, one pair at $U(t)/2$ and the other at $-U(t)/2$. The resulting oscillating field is characterized by the geometry tensor $\mathsf{A}=\bls{e}_y\otimes\bls{e}_y-\bls{e}_x\otimes\bls{e}_x$. In order to achieve confinement along all three spatial directions, two endcap electrodes with distance $\ell_{\rm ec}$ are inserted  along the ${\bf e}_z$ axis. The endcaps are held at a voltage of $U_{\rm ec}$ with respect to ground. This trap geometry was used for instance  in the experiments reported in \cite{joseph2009long,molhave2000formation,hashemloo2015rotational}.

The effective  potential \eqref{sec_2_effective_potential} produced by the rods at $U_{\rm dc}=0$ serves to trap a cylindrically symmetric particle, except for the motion in the $z$-direction. It takes the form \eqref{appendix_potential_ringtrap} with the geometry tensor of the linear Paul trap. The endcap electrodes add to Eq.~\eqref{appendix_potential_ringtrap} the potential \cite{berkeland1998minimization}
\begin{align}\label{eq:DeltaV}
\Delta V({\bf r},\Omega)= & -\frac{k_{\rm ec}U_{\rm ec}}{q\ell_{\rm ec}^2}(q\bls{r} + p_3 \blg{m} )\cdot\mathsf{A}_{\rm ec}(q\bls{r} + p_3 \blg{m} )+ \frac{k_{\rm ec}U_{\rm ec}}{q\ell_{\rm ec}^2} \left ( p_3^2 - q Q_3 \right ) {\bf m} \cdot {\sf A}_{\rm ec} {\bf m}.
\end{align}
with the geometry tensor $\mathsf{A}_{\rm ec}=\mathbb{1}-3\bls{e}_z\otimes\bls{e}_z$ and a numerical factor $k_{\rm ec}\leq 1$ describing the shape of the electrodes. 

The particle is stably trapped if the sum of the first terms of Eq.~\eqref{appendix_potential_ringtrap} and \eqref{eq:DeltaV} yields real trapping frequencies for all center-of-mass degrees of freedom. This is the case if the tensor
\begin{equation}
{\sf B} = \mathsf{A}^2-\frac{4 m k_{\rm ec}U_{\rm ec}}{q} \left ( \frac{\omega_{\rm ac}\ell_0^2}{\ell_{\rm ec}U_{\rm ac}}\right )^2 \mathsf{A}_{\rm ec},
\end{equation}
is positive definite,
\begin{equation}
0<     \frac{4 m k_{\rm ec}U_{\rm ec}}{q} \left ( \frac{\omega_{\rm ac}\ell_0^2}{\ell_{\rm ec}U_{\rm ac}}\right )^2 < 1.
\end{equation}

As in the ring-shaped trap, the center-of-mass minima are at the positions $\bls{r}=-p_3\bls{m}/q$. The second term of \eqref{appendix_potential_ringtrap} vanishes if $p_3^2=qQ_3$ or if $\bls{m}$ is parallel to $\pm\bls{e}_x$, $\pm\bls{e}_y$ or $\pm\bls{e}_z$. If the second term in Eq.~\eqref{eq:DeltaV} is positive, $U_{\rm ec}(p_3^2/q-Q_3)>0$, the particle tends to align with the $z$-axis, yielding $\bls{m}=\pm \bls{e}_z$. If it is negative
the particle aligns with the $x$- or $y$-axes, i.e.\ $\bls{m}=\pm \bls{e}_x$ or $\bls{m}=\pm \bls{e}_y$. All these equilibrium positions are plotted in Fig. \ref{figure3}(b).

\section{Equilibration in Paul traps}\label{sec:equilibration}

Having  the effective Paul trap potential at hand, we can now discuss the  thermalization to be expected if the Paul trap electrodes are connected via a parallel or series RLC circuit. In course of this, we derive the effective equilibrium state of motion  in presence of  isotropic damping and diffusion, as effected for instance by a homogeneous background gas \cite{martinetz2018}.

The fact that the micromotion momentum and angular momentum have similar magnitude as the macromotion momenta modifies the thermalized phase-space distribution. In the relevant case that the Paul trap drive frequency $\omega_{\rm ac}$ is much greater than the damping rate, friction and diffusion add only to the macromotion equations of motion, so that the macromotion approaches a Boltzmann distribution with the effective potential \eqref{sec_2_effective_potential}. Adding the micromotion momenta, yields the time-dependent phase-space distribution
\begin{align}\label{dist}
    & f_t({\bf R},{\bf P},\Omega,p_\Omega) = \frac{1}{Z}
    \exp\left[-\frac{\left(\bls{P}+\Delta\bls{P}_t\right)^2}{2mk_{\rm B}T} -\frac{V_{\rm eff}(\bls{R},\Omega)}{k_{\rm B} T} \right ]\nonumber \\
    & \times \exp \left [ - \frac{1}{2 k_{\rm B} T} \left(\bls{J}+\Delta\bls{J}_t\right)\cdot{\rm I}^{-1}(\Omega)\left(\bls{J}+\Delta\bls{J}_t\right)\right].
\end{align}
Here, $Z$ is the partition function and the function is normalised with respect to the phase space measure $d\Gamma = d^3{\bf R} d^3{\bf P} d\alpha d\beta d\gamma dp_\alpha d p_\beta d p_\gamma$. The time-dependent micromotion amplitudes $\Delta\bls{P}_t$, $\Delta\bls{J}_t$ are functions of the particle position and orientation
\begin{subequations}
\begin{align}
    \Delta\bls{P}_t&=\frac{U_{\rm ac}\sin(\omega_{\rm ac}t)}{\omega_{\rm ac}\ell_0^2}\mathsf{A}\left(q\bls{R}+\blg{p}\right),
\\
\Delta\bls{J}_t&=\frac{U_{\rm ac}\sin(\omega_{\rm ac}t)}{\omega_{\rm ac}\ell_0^2}\left(\blg{p}\times\mathsf{A}\bls{R}-\frac{1}{3}\sum_{i = 1}^3a_i\bls{a}_i\times\mathsf{Q}\bls{a}_i\right).
\end{align}
\end{subequations}
The angular momentum vector ${\bf J}$ in Eq.~\eqref{dist} is understood in terms of the rotational phase space coordinates $(\Omega,p_\Omega)$, see \ref{appendix_hamiltonian}.

The distribution \eqref{dist}  describes that the nanoparticle follows the fast Paul trap drive, giving rise to oscillating momentum variances and coordinate-momentum covariances. The resulting cycle-averaged distribution will thus in general \emph{not} have the form of a Boltzmann distribution. For instance, in the case of a particle with vanishing dipole moment, the $z$-momentum marginal is of the form
\begin{align}\label{time_dependent_distribution}
    f_{t}(P_z) = & \frac{1}{N_t} \exp\left[-\frac{1}{2m k_{\rm B}T}\frac{P_z^2}{2-\cos(2\omega_{\rm ac}t)}\right],
\end{align}
with  normalization $N_t$. This
yields the cycle-averaged kinetic energy expectation value
\begin{equation}
    \left \langle \frac{P_z^2}{2m} \right \rangle = k_{\rm B} T,
\end{equation}
which is twice the value expected in a static harmonic potential \cite{cirac1994laser}.

\begin{figure}
\centering
\includegraphics[width = 0.6\textwidth]{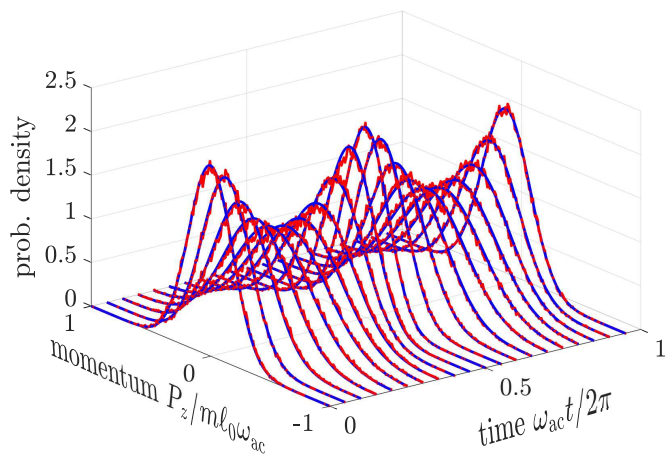}
\caption{Equilibrated center-of-mass momentum $P_z$ distribution as a function of time in the ring-shaped Paul trap for vanishing dipole moment. The full time-dependent stochastic dynamics (red) are determined by the AC voltage $qU_{\rm ac}/m\omega_{\rm ac}^2\ell_0^2=0.0034$, the environment temperature $k_{\rm B}T/m\omega_{\rm ac}^2\ell_0^2=0.034$, and the linear damping rate $\Gamma/\omega_{\rm ac}=0.02$. The analytic distribution \eqref{time_dependent_distribution} is shown in blue. 
} \label{figure4}
\end{figure}

Figure \ref{figure4} compares the momentum distribution \eqref{time_dependent_distribution} with exact numerical simulations of the stochastic $R_z = {\bf R}\cdot {\bf e}_z$ trajectories in the time-dependent Paul-trap potential, showing excellent agreement. We solved the equations of motion $dR_z=P_z dt/m$ and
\begin{equation}
    dP_z={F}_{\rm tr}(z,t)dt -\Gamma P_z dt+\sqrt{2m\Gamma k_{\rm B}T}\,dW_t,
\end{equation}
with the Wiener increment $dW_t$. In the ring-shaped Paul trap the trapping force ${F}_{\rm tr}(z,t)=\bls{F}_{\rm tr}(t)\cdot \bls{e}_z$ is independent of all other particle coordinates, see Eq.~\eqref{sec2_force}.

Finally, we remark that the coordinate marginal of the distribution \eqref{dist} is constant in time and given by the Boltzmann factor of the macromotion potential \eqref{sec_2_effective_potential},
\begin{equation}
    f({\bf R},\Omega) = \frac{\sin \beta}{Z'} \exp \left [ -\frac{V_{\rm eff}({\bf R},\Omega)}{k_{\rm B} T} \right ].
\end{equation}
where $\cos \beta = {\bf e}_z \cdot {\bf N}_3$ is the polar angle.

\section{Cooling simulation} \label{sec:exp}

All ingredients are now available to simulate the cooling dynamics of a micron-sized particle in a Paul trap. Specifically, we shall demonstrate that both the $z$-motion and the nutation of a cylindrically symmetric particle can be cooled resistively in the linear Paul trap described in Sec.~\ref{secV:linear_trap}. The endcaps are connected to a parallel RLC circuit, realising the linear pick-up configuration discussed in Sec.~\ref{main_resistive_cooling}. We will see that consecutive centre-of-mass and rotational cooling can be achieved by tuning the damping rates \eqref{sec3B_cooling_rates} by means of the circuit parameters. \tcr{The ultimate cooling limit is given by the temperature of the resistor provided that additional noise is negligible (see below). Low particle temperatures thus demand for cooling the resistor.}

\begin{figure*}
\centering
\includegraphics[width = 0.99\textwidth]{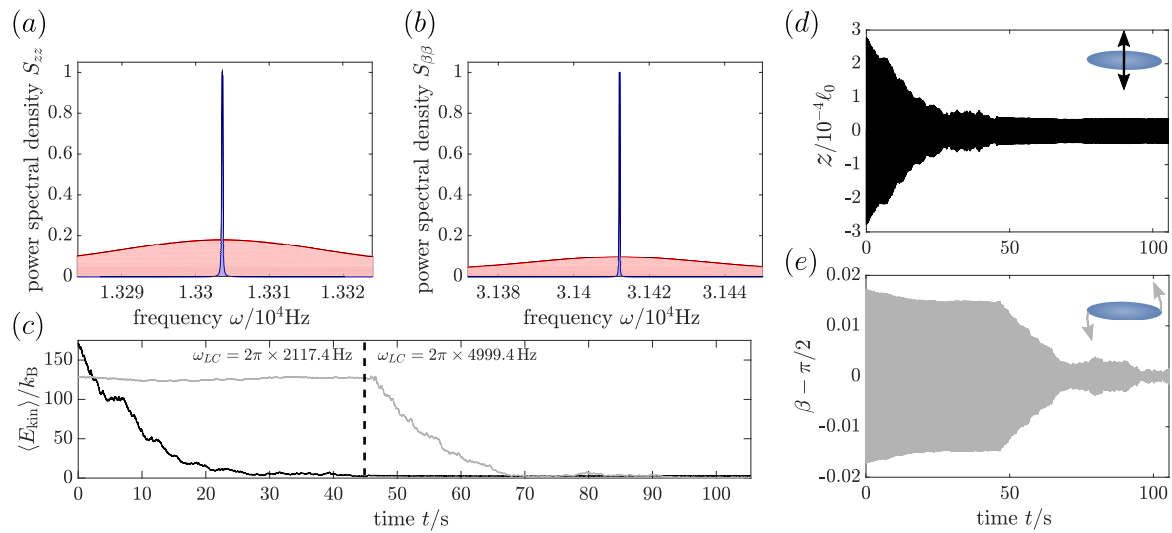}
\caption{Consecutive cooling of the center-of-mass coordinate $z=\bls{r}\cdot\bls{e}_z$ and the polar angle $\beta$ \tcr{in a cryogenic circuit} by switching the circuit parameters. Panel (a) compares the position power spectral densities (PSD) for a particle thermalized in a thin background gas at room temperature (red) and the same particle resonantly coupled to the circuit in high vacuum (blue). These analytical PSDs, given by \eqref{sec:7_PSD}, are normalized to the maximum of the blue PSD. (b) After tuning the circuit capacitance and resistance to the trapping frequency of $\beta$, the angular PSD shows cooling of the rotational motion. As in (a), the red curve displays the hot state in a thin gas while the blue curve gives the circuit-cooled PSD. Panel (c) demonstrates consecutive cooling of a stochastic trajectory. The cycle-averaged center-of-mass kinetic energy $E_{\rm kin}=p^2/2m$ (black) decreases first, while the rotational kinetic energy $E_{\rm kin}=p_\beta^2/2I_1$ (grey) remains approximately constant. After tuning the circuit, at $45$s, the rotational energy decreases while the center-of-mass energy stays constant. The amplitudes of the resulting center-of-mass and rotational trajectories are shown in (d)-(e).  The simulation parameters are given in \ref{app_simulation}.}\label{figure5}
\end{figure*}

We consider a cylindrical particle in a linear Paul trap whose dynamics at room temperature are well described by the effective potential \eqref{sec_2_effective_potential}. If the particle motion is confined to the region close to the potential minimum at $x=z=0$, $y=-p_3/q$ and $\beta=\alpha=\pi/2$ (see Fig.~\ref{figure3}), the harmonic approximation is justified. The two degrees of freedom $\beta$ and $z$ are then coupled linearly in the effective potential \eqref{sec_2_effective_potential}, giving rise to two normal modes which can be associated with $z$ and $\beta$ for small coupling.

In this approximation the dynamics are described by a first order linear differential equation for the tuple $\boldsymbol{\xi}=(z,\beta-\pi/2,Q, p, p_\beta, \Phi )$,
\begin{equation}\label{sec:7_linear_dgl}
    d\boldsymbol{\xi}=\mathsf{B}\boldsymbol{\xi}dt +\mathsf{N}d\boldsymbol{W}_t.
\end{equation}
Here, ${\sf B}$ describes the deterministic dynamics and ${\sf N}d\boldsymbol{W}_t$ accounts for the noise. The components of the tuple of Wiener increments $d\boldsymbol{W}_t$ are taken to be uncorrelated.

For a linear Paul trap connected to a parallel RLC circuit, the coupling of all degrees of freedom and the damping due to the circuit and the background gas is described by the matrix 
\begin{equation}
    \mathsf{B}=\left( \begin{array}{cccccc}
0 & 0 & 0 & 1/m & 0 & 0 \\ 
0 & 0 & 0 & 0 & 1/I_1 & 0\\
-g_{zQ}/R & -g_{\beta Q}/R & -1/RC & 0 & 0 & 1/L\\ 
-m\omega_{z}^2 & -g_{z\beta} & -g_{zQ} & -\Gamma_z & 0 & 0 \\ 
-g_{z\beta} & -I_1\omega_{\beta}^2 & -g_{\beta Q} & 0 & -\Gamma_\beta & 0\\
-g_{zQ} & -g_{\beta Q} & -L\omega_{LC}^2 & 0 & 0 & 0\\ 
\end{array}\right),
\end{equation}
with the coupling constants $g_{zQ}=kq/Cz_0$, $g_{\beta Q}=-kp_3/Cz_0$
and
\begin{equation}
    g_{z\beta}=-\frac{4k_{\rm ec}U_{\rm ec}p_3}{\ell_{\rm ec}^2}-\frac{k^2qp_3}{Cz_0^2}.
\end{equation}
The harmonic frequencies of the uncoupled mechanical modes are
\begin{subequations}
\begin{align}
\omega_z^2=&\frac{4k_{\rm ec}U_{\rm ec}q}{m\ell_{\rm ec}^2}+\frac{k^2q^2}{m Cz_0^2}
\intertext{and}
\omega_{\beta}^2=&\frac{2k_{\rm ec}U_{\rm ec}}{I_1 \ell_{\rm ec}^2}\left(3Q_3-\frac{p_3^2}{q}\right)+\frac{k^2p_3^2}{I_1C z_0^2}+\frac{U_{\rm ac}^2}{2I^2_1\ell_0^4\omega_{\rm ac}^2}\left(Q_3-\frac{p_3^2}{q}\right)^2 .
\end{align}
\end{subequations}
The mechanical gas damping rates $\Gamma_z$ and $\Gamma_\beta$, calculated as described in Ref.\ \cite{martinetz2018}, are accompanied with noise according to the fluctuation-dissipation theorem. The magnitude of this noise is described by the matrix
\begin{equation}
    \mathsf{N}=\left( \begin{array}{cccccc}
0 & 0 & 0 & 0 & 0 & 0 \\ 
0 & 0 & 0 & 0 & 0 & 0\\
0 & 0 & \sqrt{2D_{\rm cir}} & 0 & 0 & 0\\ 
0 & 0 & 0 & \sqrt{2D_z} & 0 & 0 \\ 
0 & 0 & 0 & 0 & \sqrt{2D_\beta} & 0\\
0 & 0 & 0 & 0 & 0 & 0\\ 
\end{array}\right)
\end{equation}
with $D_{\rm cir}=C\tcr{\gamma_p}k_{\rm B}T_{\rm cir}$, $D_{z}=m\Gamma_zk_{\rm B}T_{\rm gas}$ and $D_{\beta}=I_1\Gamma_\beta k_{\rm B}T_{\rm gas}$. \tcr{Noise due to fluctuating currents in the electrode material can safely be neglected for common highly conductive electrode materials and the electrode-particle distance in our simulation \cite{kumph2016electric}.}

The steady-state power spectral density follows from Eq.\,\eqref{sec:7_linear_dgl} as the Fourier transform of the correlation matrix,
\begin{equation}\label{sec:7_PSD}
    \mathsf{S}(\omega)=\frac{1}{2\pi}\left(i\omega\mathbb{1}-\mathsf{B}\right)^{-1}\mathsf{N}\mathsf{N}^{\rm T} \left[\left(-i\omega\mathbb{1}-\mathsf{B}\right)^{-1}\right]^{\rm T}.
\end{equation}
Its diagonal elements yield the power spectral densities of the individual degrees of freedom, as plotted in Fig.~\ref{figure5}(a), (b).\tcr{The area under each power spectral density determines the final effective temperature of the corresponding degree of freedom.}

Figure \ref{figure5} displays consecutive translational and rotational cooling of a deeply trapped particle with a parallel RLC circuit. The circuit resistance and capacitance are changed during the cooling process after $45$ seconds. In Fig.~\ref{figure5}(a) we compare the steady-state position PSD from equation (\ref{sec:7_PSD}) for an off-resonant circuit (red) with the corresponding PSD (blue) for a circuit in resonance with the center-of-mass frequency. In Fig.~\ref{figure5}(b) we compare the steady-state libration PSDs of the off-resonant case (red) with the corresponding PSD  for the circuit in resonance with the libration frequency  (blue). Both panels show significant mechanical cooling due to dissipation through the circuit. The center-of-mass and the rotational kinetic energy in Fig.~\ref{figure5}(c) for one stochastic trajectory (e-d) demonstrate, that position and orientation can be separately cooled. Here, the circuit resonance frequency is changed at $45\,{\rm s}$ from cooling the center-of-mass to the rotational motion. The figure shows that this setup can realistically achieve temperatures on the order of a few Kelvin on a timescale of seconds.

\section{Discussion} \label{sec:conclusions}

In conclusion, we showed how the rotational and translational motion of levitated charged  nanoparticles can be controlled with electric circuitry. We derived the effective trapping potential for aspherical objects with non-vanishing permanent electric multipole moments in quadrupole ion traps, and numerically demonstrated cooling in a realistic setup. The derived relations will be relevant for quantum experiments with charged  molecules and nanoparticles \cite{joseph2009long,molhave2000formation,hashemloo2015rotational,delord2017a,delord2018,martinetz2020}.

The methods discussed in this article can also be used to trap and control metallic nanoparticles \cite{chen2015fano,lopez2018internal,roda2020quantum}, whose large absorption cross section precludes optical trapping. However, in contrast to non-conducting objects, metallic particles can have significant induced electric moments, which can contribute to the dynamics in the trap and to the coupling to electric circuits. Whether induced moments play a role depends strongly on the particle size, shape, total charge, and trapping field. 

For instance, the charge distribution on the surface of a metallic spheroid of total charge $q$ in the absence of external fields is described by the surface charge density
\begin{equation}\label{metallic_particles_distribution_static_dipole}
\sigma=\frac{q}{4\pi a r^2}\left(\frac{x^2+y^2}{r^4}+\frac{z^2}{a^4}\right)^{-1/2}.
\end{equation}
Here, $r$ denotes the radius and $2a$ the length of the spheroid. This charge distribution has vanishing dipole moment, while the quadrupole tensor reads
\begin{equation}\label{eq:quadrmet}
    \mathsf{Q}=\frac{qd^2}{3}\left(3\bls{m}\otimes\bls{m}-\mathbb{1}\right),
\end{equation}
where ${\bf m}$ is the spheroid main axis and $d^2 = a^2 - r^2$.

The induced quadrupole moment can be readily obtained for a nearly spherical particle. In the center of a Paul trap it is given by
\begin{equation}\label{indquadr}
    \mathsf{Q}_{\rm ind}=-4\pi \epsilon_0 U(t) \frac{ a^5 }{\ell_0^2} \mathsf{A}.
\end{equation}
For highly charged, aspherical particles and moderate trapping voltages, this can be neglected compared to the much greater permanent quadrupole moments \eqref{eq:quadrmet}. In contrast, the induced dipole moment depends on the field strength $E$ at the centre-of-mass position. For prolate particles $a > r$, the dipole moment is maximal if the electric field is aligned with the particle main axis. The corresponding polarizability $\alpha_{\rm max} = p_{\rm max}/E$  follows as
\begin{equation}\label{inddip}
\alpha_{\rm max}=\frac{4\pi\epsilon_0 d^3}{3}\left [\ln\left(\frac{a+d}{r}\right)-\frac{d}{a}\right ]^{-1}.
\end{equation} 
The induced dipole moment dominates for perfectly inversion symmetric particles with no permanent dipole moments. However, for realistic situations, the centre-of-charge and the centre-of-mass do not coincide. For instance, a particle of the shape of two joined half-spheroids of lengths $a$ and $a + \Delta a$ gives rise to the approximate permanent dipole moment $\blg{p}=q\Delta a/8$, which dominates for highly charged microscale particles at deviations as small as a few percent and moderate trapping voltages.

If the induced moments are relevant, the effective potential \eqref{sec_2_effective_potential} for a rigid charge distribution must be generalized accordingly. Knowing the induced dipole and quadrupole moments, which are in general complicated to calculate analytically, one can derive the corresponding effective potential, cooling rates, and coupled nanoparticle-circuit dynamics by using the methodology developed in this work.

\section*{Acknowledgements}
We thank Tobias Kuhn for discussions. This research was funded by the Deutsche Forschungsgemeinschaft (DFG, German Research Foundation) -- 411042854.

\appendix

\section{Rotational phase-space coordinates and alternative Hamiltonian}\label{appendix_hamiltonian}

Parametrizing the orientation of the nanoparticle by the Euler angles $\Omega=(\alpha, \beta, \gamma)$  in the $z$-$y'$-$z''$ convention \cite{sakurai,timothesis}, the body-fixed axes take the form
\begin{subequations}
\begin{align}
    {\bf N}_1 = &\, (\cos \alpha \cos \beta \cos \gamma - \sin \alpha \sin \gamma) {\bf e}_x+ (\sin \alpha \cos \beta \cos \gamma + \cos \alpha \sin \gamma ) {\bf e}_y \nonumber \\
    & -\sin \beta \cos \gamma {\bf e}_z \\
    {\bf N}_2 = &\, (-\cos \alpha \cos \beta \sin \gamma - \sin \alpha \cos \gamma) {\bf e}_x+ (-\sin \alpha \cos \beta \sin \gamma + \cos \alpha \cos \gamma ) {\bf e}_y \nonumber \\
    & +\sin \beta \sin \gamma {\bf e}_z \\
    {\bf N}_3 = &\, \cos \alpha \sin \beta {\bf e}_x+ \sin \alpha \sin \beta {\bf e}_y+\cos \beta {\bf e}_z .
\end{align}
\end{subequations}

The conjugate angular momenta $\blg{p}_\Omega=(p_\alpha,p_\beta,p_\gamma)$ are related to the angular momentum vector $\bls{J}$ by $p_\alpha=\bls{J}\cdot \bls{e}_z$, $p_\beta=\bls{J}\cdot \bls{e}_\xi$ and $p_\gamma=\bls{J}\cdot \bls{N}_3$. Here $\bls{e}_\xi = -\sin \alpha {\bf e}_x + \cos\alpha {\bf e}_y$ denotes the nodal line. The body-fixed angular momentum components then follow as
\begin{subequations}
\begin{align}
\bls{J}\cdot\bls{N}_1&=-\frac{\cos\gamma}{\sin\beta}p_\alpha+\sin\gamma p_\beta+\cot\beta\cos\gamma p_\gamma, \\
\bls{J}\cdot\bls{N}_2&=\frac{\sin\gamma}{\sin\beta}p_\alpha+\cos\gamma p_\beta-\cot\beta\sin\gamma p_\gamma,\\
\bls{J}\cdot\bls{N}_3&=p_\gamma.
\end{align}
\end{subequations}

The Hamiltonian (\ref{sec3_hamiltonian}) can be rewritten with a canonical transformation by introducing the capacitor charge $Q'=CU_z=Q+Q_{\rm ind}(\bls{R},\Omega)$ as the circuit coordinate with associated conjugate momentum $\Phi$. While the particle coordinates are not affected, the transformation changes the canonical particle momenta to $P_i'= P_i-\Phi\partial Q_{\rm ind}/\partial R_i$, and $p_\mu' = p_\mu-\Phi\partial Q_{\rm ind}/\partial\mu$, where $\mu =\alpha,\beta,\gamma$, which are no longer equal to the kinetic linear and angular momentum. The resulting Hamiltonian reads
\begin{align}\label{appendix_new_hamiltonian}
H'=&\frac{1}{2m}\left({\bls{P}}'+\Phi\frac{\partial Q_{\rm ind}}{\partial {\bf R}}\right )^2  
+\frac{\Phi^2}{2L}+\frac{Q'^2}{2C}+V_{\rm tr}(\bls{R},\Omega,t)+V_{\rm im}(\bls{R},\Omega)\nonumber \\
& + \frac{1}{2}\left({\bls{J}'}+\Phi{\bf T}\right)\cdot{\rm I}^{-1}\left({\bls{J}}'+\Phi {\bf T}\right).
\end{align}
Here the dependence between $\bls{J}'$ and the $p_\mu'$ is defined in analogy to the relation between $\bls{J}$ and the $p_\mu$ or $\bls{T}$ and $\partial Q_{\rm ind}/\partial \mu$.

The Hamiltonian \eqref{appendix_new_hamiltonian} can be  useful when describing the interaction with a parallel RLC circuit. In this case, the non-conservative charge dynamics take the simple form
\begin{equation}
\dot Q'=\frac{\partial H'}{\partial \Phi}-\frac{1}{RC}Q'+I_{\rm fl}(t),
\end{equation}
showing that the capacitor charge and voltage decay due to the resistance in the circuit. The dynamics of all other phase space coordinates follow from \eqref{appendix_new_hamiltonian} with Hamilton's equations.

\section{Image force and torque}\label{appendix_image_potential}

The image force and torque experienced by a charged particle between two infinitely extended, parallel metal plates with distance $2z_0$ follows from repeatedly applying the method of image charges. In particular, the point charge $q$ at position $\bls{r}_{0}$ with respect to the capacitor center induces an infinite sequence of positive and negative image charges. The positive charges $q$ are located at $\bls{r}_{\rm ic}=\bls{r}_{\rm 0}\pm 4nz_0\bls{e}_z$, where $n\in \mathbb{N}$ and ${\bf e}_z$ is the normal direction of the plates. The negative image charges $-q$ are positioned at $\mathsf{M}\bls{r}_{\rm 0}\pm (4n-2)z_0\bls{e}_z$ with $\mathsf{M}=(\mathbb{1}-2\bls{e}_z\otimes\bls{e}_z)$.

The $n$-th pair of positive image charges gives rise to the electric field
\begin{equation}
    \bls{E}_n^{(+)}({\bf r}) \simeq -\frac{q}{2\pi\varepsilon_0}\frac{1}{(4nz_0)^3}(3\bls{e}_z\otimes\bls{e}_z-\mathbb{1})(\bls{r}-\bls{r}_{\rm 0}),
\end{equation}
at positions ${\bf r}$ close to the centre, $|\bls{r}-\bls{r}_{\rm 0}|\ll z_0$. Likewise, the $n$-pair of negative image charges induces
\begin{equation}
    \bls{E}_n^{(-)}({\bf r}) \simeq \frac{q}{2\pi\varepsilon_0}\frac{1}{[(4n-2)z_0]^3}(3\bls{e}_z\otimes\bls{e}_z-\mathbb{1})\left (\bls{r}-\mathsf{M}\bls{r}_{\rm 0}\right).
\end{equation}
Summing over all image charges and integrating the resulting field over the charge distribution of the levitated particle, yields the total electric field due to the image charges
\begin{align}\label{app_total_electr_field_image}
    \bls{E}(\bls{r})\simeq&\frac{\zeta(3)}{64\pi\varepsilon_0z_0^3}(3\bls{e}_z\otimes\bls{e}_z-\mathbb{1})\left[3(q\bls{r}-q\bls{R}-\blg{p} ) +7 {\bf e}_z \,[{\bf e}_z \cdot (q\bls{R}+\blg{p})]\right],\nonumber\\
\end{align}
with $\zeta(\cdot)$ the Riemann $\zeta$-function. The resulting force and torque (\ref{image_force_torque}) follow from integrating the electric field (\ref{app_total_electr_field_image}) with the particle charge distribution.

\begin{figure}
\centering
\includegraphics[width = 0.6\textwidth]{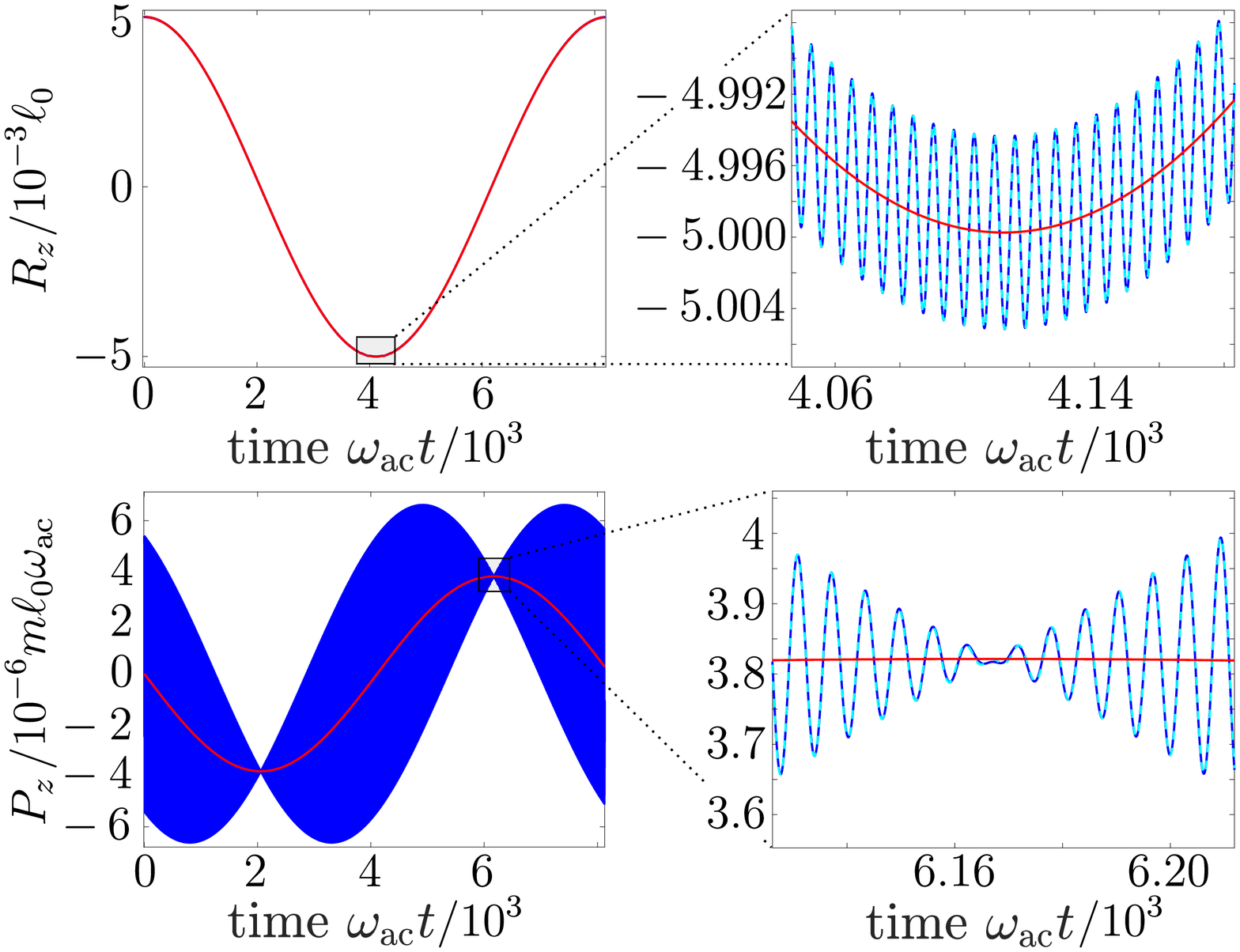}
\caption{In the limit of high drive frequencies, the center of mass performs an approximately harmonic oscillation. The exact trajectory and momentum oscillations (dark blue solid line) are very well described by the macromotion dynamics (red solid line) if the momentum correction (\ref{korrektur_mom_und_ang_mom}) is included (light blue dashed line).} \label{figure6}
\end{figure}

\section{Simulation parameters}\label{app_simulation}

Figure \ref{figure2} shows the dynamics of an asymmetric $10^6$\,amu silicon particle with moments of inertia $I_1=I_0$, $I_2=0.92I_0$, $I_3=0.55I_0$ and $I_0=2.8\times 10^{-38}\,{\rm kg\, m^2}$. Its charge distribution is characterized by the total charge $q=200e$, the body-fixed dipole moments $p_1=0.0025q\ell$, $p_2=0.0022q\ell$ and $p_3=0.007q\ell$ and the body-fixed quadrupole moments $Q_{11}=-0.13q\ell^2$, $Q_{12}=0.08q\ell^2$, $Q_{13}=0.24q\ell^2$, $Q_{22}=-0.04q\ell^2$ and $Q_{23}=0.03q\ell^2$, which depend on the particle length scale $\ell=12$\,nm. The trapping field of the ring-shaped Paul trap is characterized by  $U_{\rm ac}=750$\,V, $U_{\rm dc}=0$, $\omega_{\rm ac}=2\pi\times 75$\,MHz and $\ell_0=0.25\sqrt{2}$\,mm. The resulting centre-of-mass trajectory is shown in Fig.~\ref{figure6}.

Figure \ref{figure5} displays consecutive translational and rotational cooling of a deeply trapped particle with a parallel RLC circuit. The circuit parameters $k=0.4$, $L=0.565\,{\rm H}$ and $T_{\rm cir}=4\,{\rm K}$ and the gas temperature $T_{\rm gas}=300\,{\rm K}$ are kept constant, the resistance and capacitance are $C_0=5.8\,{\rm nF}$, $R_0=2\,{\rm M\Omega}$, and $p_{\rm g} = 0.1\,{\rm mbar}$ initially, $R_{\rm cm}=2\,{\rm M\Omega}$, $C_{\rm cm}=10\,{\rm nF}$, $p_{\rm g} = 10^{-8}\,{\rm mbar}$ for center-of-mass cooling and $R_{\rm rot}=11.15\,{\rm M\Omega}$, $C_{\rm rot}=1.794\,{\rm nF}$, and $p_{\rm g} = 10^{-8}\,{\rm mbar}$ for rotational cooling. The particle and circuit start with $z=0$, $\beta=\pi/2$, $p=-\sqrt{2mk_{\rm B}T_{\rm gas}}$, $p_\beta=-\sqrt{2I_1k_{\rm B}T_{\rm gas}}$, $Q=-4.8\,e$ and $\Phi=0.4\,e L \omega_{\rm ec}$. The circuit resonance frequency  changes at $45\,{\rm s}$ from $1/\sqrt{LC_{\rm cm}}=2\pi\times 2117.4\,{\rm Hz}$ to $1/\sqrt{LC_{\rm rot}}=2\pi\times 4999.4\,{\rm Hz}$. The trapping potential is specified by $U_{\rm ac}=5000\,{\rm V}$, $\omega_{\rm ac}/2\pi = 750\,{\rm kHz}$, $\ell_0/\sqrt{2}=250\,{\rm \mu m}$, $\ell_{\rm ec}=2z_0=2\ell_0$ and $k_{\rm ec}=1$. The freely floating endcap electrodes are on a voltage of $U_{\rm ec}=8.24\,{\rm V}$ with respect to ground. The cylindrically symmetric silicon particle has a total charge of $q=10^5\,e$ and a length of $\ell=2500\,{\rm nm}$, so that $p_3=0.1q\ell$, $Q_3=0.15q\ell^2$, $m=3.5\times 10^{12}\,{\rm amu}$, $I_1=3.52\times10^{-27}\,{\rm kg\,m^2}$ and $I_2=I_1$. The gas damping rates at $p_{\rm g} = 0.1\,$mbar are $\Gamma_z=44.5\,{\rm Hz}$ and $\Gamma_\beta=77.8\,{\rm Hz}$ and at $p_{\rm g} = 10^{-8}\,$mbar are $\Gamma_z=4.5\times10^{-6}\,{\rm Hz}$ and $\Gamma_\beta=7.8\times10^{-6}\,{\rm Hz}$ \cite{martinetz2018}.

\newpage
\providecommand{\newblock}{}

\end{document}